\documentclass[twocolumn]{aastex701}

\usepackage{amsmath}
\usepackage{multirow}
\usepackage{listings}

\usepackage{xeCJK}
\setCJKsansfont{ipam.ttf}
\setCJKmonofont{ipam.ttf}

\newcommand{\msun}{M$_\odot$}

\newcommand{\hi}{\ion{H}{1}}
\newcommand{\hii}{\ion{H}{2}}
\newcommand{\mgii}{\ion{Mg}{2}}
\newcommand{\ovi}{\ion{O}{6}}

\newcommand{\trident}{\textsc{trident}}

 \newcommand{\rev}[2]{#2}

\makeatletter 
\patchcmd{\NAT@citex}
{\@citea\NAT@hyper@{%
		\NAT@nmfmt{\NAT@nm}%
		\hyper@natlinkbreak{\NAT@aysep\NAT@spacechar}{\@citeb\@extra@b@citeb}%
		\NAT@date}}
{\@citea\NAT@nmfmt{\NAT@nm}%
	\NAT@aysep\NAT@spacechar\NAT@hyper@{\NAT@date}}{}{}

\patchcmd{\NAT@citex}
{\@citea\NAT@hyper@{%
		\NAT@nmfmt{\NAT@nm}%
		\hyper@natlinkbreak{\NAT@spacechar\NAT@@open\if*#1*\else#1\NAT@spacechar\fi}%
		{\@citeb\@extra@b@citeb}%
		\NAT@date}}
{\@citea\NAT@nmfmt{\NAT@nm}%
	\NAT@spacechar\NAT@@open\if*#1*\else#1\NAT@spacechar\fi\NAT@hyper@{\NAT@date}}
{}{}
\makeatother

\begin{document}
	
	\title{ENhanced Galactic Atmospheres With Arepo:\\Resolving the CGM at 200~pc with the ENGAWA Simulations}
	
	\author[0000-0001-9982-0241]{Scott Lucchini}
	\affiliation{Center for Astrophysics $|$ Harvard \& Smithsonian, 60 Garden Street, Cambridge, MA 02138, USA}
	\email{scott.lucchini@cfa.harvard.edu}
	
	\correspondingauthor{Scott Lucchini}
	\email{scott.lucchini@cfa.harvard.edu}
	
	\author[0009-0004-2769-621X]{Cecilia Abramson}
	\affiliation{California Institute of Technology, Pasadena, CA 91125, USA}
	\email{cecilia@caltech.edu}
	
	\author[0000-0002-3817-8133]{Cameron Hummels}
	\affiliation{California Institute of Technology, Pasadena, CA 91125, USA}
	\email{chummels@caltech.edu}
	
	\author[0000-0002-1590-8551]{Charlie Conroy}
	\affiliation{Center for Astrophysics $|$ Harvard \& Smithsonian, 60 Garden Street, Cambridge, MA 02138, USA}
	\email{cconroy@cfa.harvard.edu}
	
	\author[0000-0001-6950-1629]{Lars Hernquist}
	\affiliation{Center for Astrophysics $|$ Harvard \& Smithsonian, 60 Garden Street, Cambridge, MA 02138, USA}
	\email{lhernquist@cfa.harvard.edu}
	
	\author[0000-0002-2838-9033]{Aaron Smith}
	\affiliation{Department of Physics, The University of Texas at Dallas, Richardson, TX, 75080, USA}
	\email{}
	
	\begin{abstract}
		
		Simulating the small-scale features and dynamics of the circumgalactic medium (CGM) is computationally challenging due to its large volume, low densities, multiphase structure, and chaotic environmental effects.
		Traditional mass-based refinement schemes focus computational power on the high-density regions, thus alternative techniques are required to study the details of the CGM.
		In this paper, we introduce a new suite of four cosmological zoom-in simulations of Milky Way-like galaxies in which we include fixed-volume refinement throughout the CGM combined with the IllustrisTNG stellar and AGN feedback model down to redshift zero. Reaching spatial resolutions of 200~pc, we see enhancements in low ion column densities (\hi\ and \mgii) and the number of cold clouds around galaxies, relieving some of the longstanding tensions between simulations and observations of the CGM.
		We additionally apply the COLT radiative transfer code in post-processing to account for stellar radiation, providing a more realistic gauge of ion populations. We find a reduction in the \hi\ with minimal impact to the \mgii\ and \ovi, tempering the impact of resolution while still providing results consistent with observations.
		In addition to the increase in the number of cold clouds in the CGM, we find that their intermediate temperature boundary regions are reduced in size as the resolution is increased,
		leading to smoother transitions to the ambient CGM temperature.
		This paper outlines initial results from this fixed-volume simulation suite
		which will serve as a basis for future explorations of CGM dynamics, gas accretion, and galaxy evolution.
		
	\end{abstract}
	
	\keywords{\uat{Magnetohydrodynamical simulations}{1966} --- \uat{Circumgalactic medium}{1879} --- \uat{Galaxies}{573}}

	\section{Introduction}
	
	Galaxy evolution at low redshift is regulated by the circumgalactic medium (CGM). For galaxies to grow, gas and dark matter must accrete from intergalactic space or smaller satellite galaxies. But in order to form new stars, the gaseous material must pass through the CGM and settle into the galactic disk, cool, and condense. The properties of the CGM directly dictate the dynamics of this process and thus the evolution and growth of galaxies \citep{putman12,tumlinson17}.
	
	
	Unfortunately, the CGM is very difficult to directly observe. Absorption line spectroscopy is able to provide us with information on the ionic composition of the gas around galaxies, but we are often limited to a single sightline per galaxy due to the lack of bright background objects. Current work has expanded our understanding of a few local galaxies with many sightlines \citep{dk22,mishra24,amiga}, however this remains possible for only a handful of systems. These broader, multi-galaxy absorption studies include the COS-Halos, COS-GASS, and DIISC programs \citep{werk14,borthakur15,gim21}, as well as the Cosmic Ultraviolet Baryon Survey (CUBS; \citealt{chen20}) and the Keck Baryonic Structure Survey (KBSS; \citealt{rudie12}) at higher redshift. They have revealed gaseous material at a large variety of temperatures and compositions in absorption out into the far reaches of galactic halos. In Milky Way-mass and smaller systems (from the COS-Halos survey), the dominant component of the CGM by mass has been shown to be the cool phase (traced by \ion{C}{2}, \ion{C}{3}, \ion{Si}{2}, and \ion{Si}{3}) bringing the total baryonic mass in the galaxy up to the cosmological fraction \citep{werk14}. Furthermore, the majority of metals produced through stellar evolution in L* galaxies are expelled from the disk and reach out to 150~kpc as traced by \ovi\ and lower ions \citep{peeples14}.
	
	\rev{}{Recently, observing the CGM in emission has also become feasible \citep{zhang16,zhang18,nielsen24}. Stacking results show a significant cool gas reservoir that could cool and fuel star formation \citep{zhang21}, in addition to extended hot halos traced in X-rays \citep{zhang24}. On the other hand, exceptional individual sources show large winds and multiphase structure in the inner circumgalactic gas \citep{hayes16,nielsen24,ha25}.}
	
	
	Simulations have long been an excellent tool to understand and explain these observations, however simulating the CGM of galaxies is also a notoriously difficult problem \citep{faucher-giguere23}. Due to low densities and immense volume, it is a computational challenge to be able to resolve small-scale structures in the CGM. This is because most simulations use gas elements with fixed mass. \rev{This is intrinsically true for Lagrangian codes (like smoothed particle hydrodynamics, SPH, e.g., GADGET, \citealt{gadget}, Gasoline, \citealt{gasoline}, SWIFT, \citealt{swift}; or modern meshless schemes, e.g., GIZMO, \citealt{gizmo}) since the gas particles cannot exchange mass with their neighbors. However, even hybrid and adaptive mesh refinement (AMR) codes use the gas cell masses and densities as the criterion for refinement and derefinement. While AMR codes use Eulerian grids, they will increase the number of cells in regions with higher densities (thus attempting to maintain a fixed amount of mass in each cell; e.g., ENZO, \citealt{enzo}, RAMSES, \citealt{ramses}). Similarly, hybrid, moving-mesh codes will refine or derefine individual cells if their masses become too large or small (e.g., Arepo, \citealt{springel10,arepo}).}{This leads to improved resolution in high density regions (i.e. the galaxies themselves) while not ``wasting'' computational effort on low density regions such as voids which historically have not provided as much scientific value. However, with this refinement technique, intermediate density regions, such as the CGM, also remain relatively poorly resolved. Furthermore, even with more computational power, simply improving the base mass resolution of the simulations is not a viable solution since} \rev{Moreover, }{}many of the star formation and stellar feedback prescriptions are calibrated for specific resolutions\rev{and}{. Parameter studies have shown that dramatically} changing the resolution of the interstellar medium (ISM) can move galaxies far from the stellar mass$-$halo mass relation and requires recalibration \citep{camels}.
	
	\rev{Thus, there are two viable paths forward: (1) increase the mass resolution of the CGM while retaining a fiducial mass resolution for the disk and inner galaxy; or (2) refine the low density regions more than the high density regions by fixing the cell volumes.}{Thus, we wish to refine the CGM while retaining a fiducial resolution for the ISM. This leads to two viable paths forward: (1) increase the mass resolution while introducing a CGM/galaxy boundary inside which the resolution does not increase; or (2) increase the spatial resolution everywhere up to the point at which the cell sizes in the CGM match those in the ISM.}
	The first of these techniques has been employed in the GIBLE simulation suite \citep{suresh19,gible} in which very low cell masses ($1.8\times10^3$~\msun) are used for the region $0.15\ R_\mathrm{200}<r<1.0\ R_\mathrm{200}$, where $R_\mathrm{200}$ measures the virial radius, starting from the seeding of the galaxy's central black hole (BH; $z\sim3-5$). Thus the galactic disk is left at relatively low resolution, while the CGM is highly refined. This technique works well for resolving the cool, dense phase in the CGM since it will continue to refine this gas as the target gas mass goes down. However, the inner CGM and disk are not refined, so interactions between these two regions are poorly modeled. 
	Furthermore, even \rev{}{refining to} very small cell masses still results in modest spatial resolution in the hot, volume-filling phase of the CGM due to its \rev{extremely }{}low densities.
	
	The second CGM-refinement technique has been employed in several previous studies including the \textsc{tempest} and FOGGIE simulations run with ENZO \citep{hummels19,foggie}, as well as the SURGE simulations done with Arepo \citep{vandevoort19}. In addition to the default mass-refinement, these simulations apply an additional volume criterion around the galaxies. \rev{Due to \textsc{tempest} and FOGGIE's use of the AMR code ENZO, they are able to force refinement around the galaxy from $z=6$.}{} FOGGIE \rev{uses level 9 corresponding to resolution elements of 1.1~ckpc}{has a maximum cell size of 1.1~ckpc at $z=0$}, extending out to $\pm100\ \mathrm{ckpc}/h$, and \textsc{tempest} \rev{uses level 10 (545~cpc) with multiple forced refinement boxes gradually reducing the refinement level}{obtains 545~cpc resolution within $\pm100\ \mathrm{ckpc}/h$ with gradually coarsening resolution} out to $\pm800\ \mathrm{ckpc}$. \rev{}{These additional refinement criteria are turned on at $z=4$ and $z=3$ for FOGGIE and \textsc{tempest}, respectively.} \rev{FOGGIE and \textsc{tempest}}{Both of these simulations} use oversimplified thermal dump feedback prescriptions which leads to overly massive stellar disks at $z=0$. The SURGE simulations use Arepo (and the Auriga feedback model) and refine all SUBFIND detected subhalos with $M_\mathrm{halo}>10^{8.7}$~\msun\ out to 1.2$R_\mathrm{200}$. The additional refinement criterion fixes the spatial resolution for cells below a certain cutoff density, reducing the size of cells to 1~kpc.
	
	\rev{}{In addition to these existing suites that reach redshift zero (and with which we are directly comparing), there are several additional CGM-refinement simulations run at higher redshift.} Recently, the MEGATRON simulations (run with RAMSES) have also included additional refinement criteria throughout the CGM \citep{cadiou25}. These advanced simulations include non-equilibrium \rev{cooling}{thermochemistry} and radiation transport, as well as refinement on the jeans length and cooling length to enhance the CGM resolution. \rev{Unfortunately}{While they reach 120~pc resolution in the CGM}, due to the increased computational cost of these features, these simulations are only run down to \rev{$z\sim 4$}{$z\sim 3$}. Additionally, the CGOLS (Cholla Galactic Outflow Simulation; \citealt{schneider18}) suite includes incredibly high, fixed spatial resolution (5~pc), however this is only possible for isolated galaxy simulations in moderate boxes ($10\times10\times20$~kpc) that only cover the \rev{edge}{very inner portion of} of the CGM.
	
	These simulations have shown that several features of simulated galaxies actually converge well at moderate mass resolution, such as the total amount of mass in the CGM \citep{foggie,gible}, the covering fractions of many ions (total H, \ion{Si}{2}, \ion{C}{4}, \ovi; \citealt{foggie,vandevoort19,gible}), and the distribution and properties of the largest clouds \citep{gible}. However, increased resolution dramatically changes the morphologies and structures within the CGM on small scales. There are many more smaller clouds with much more well-sampled velocity structure \citep{hummels19,gible,augustin25}. Due to the increased resolved turbulence, hydrostatic and virial equilibria do not well describe galactic atmospheres \citep{foggie5,foggie6}. Additionally, because the density and temperature structure in each cloud is better resolved, the higher resolution simulations see more cool and cold gas condensing out of the CGM \citep{hummels19,vandevoort19}\rev{}{, in addition to changes in the ionization state with the CGM possibly out of photoionization equilibrium \citep{cadiou25}}. This leads to increases in \hi\ covering fractions, however the magnitude of this increase seems to be dependent on the galaxies' histories or on the simulation code used \citep{gible}.
	
	\begin{deluxetable*}{lccccc}
		\tablecaption{Properties of selected galaxies at $z=0$}
		\label{tab:galaxies}
		\tablehead{ \colhead{Galaxy Name} & \colhead{$M_*$} & \colhead{$M_{200}$} & \colhead{$R_{200}$} & \colhead{Max resolution} \\
			& ($10^{10}$~\msun) & ($10^{10}$~\msun) & (kpc) & (pc)}
		
		\startdata
		Au6 & 4.8 & 104.4 & 214 & 200 \\
		Au8 & 3.0 & 108.1 & 216 & 500 \\
		TNG-A (537941) & 5.2 & 102.2 & 212 & 200 \\
		TNG-B (519311) & 9.2 & 139.4 & 236 &  500
		\enddata
		
		\tablecomments{Stellar and virial masses for the Auriga galaxies (Au6 and Au8) were taken from \citet{grand17}. For the TNG galaxies, the values were taken from the online halo catalog for these subhalos.\vspace{-0.8cm}}
	\end{deluxetable*}
	
	To understand how these simulation techniques relate to the observational mysteries stated above, we introduce the 
	ENhanced Galactic Atmospheres With Arepo, or ENGAWA simulations in which we specifically focus on resolving and studying the inner circumgalactic medium and disk-halo interface$-$the gateway between a galaxy and its surroundings\footnote{The name originates from the Japanese architectural feature of the covered wooden veranda, or \textit{engawa} (縁側), which serves as the gateway between the home and the surrounding nature.}. These simulations feature enhanced resolution in the CGM (with fixed maximum gas cell sizes) and improve upon previous works by pushing to higher resolution \rev{}{at redshift zero} (200 pc spatial scales) and continuously maintaining this level of resolution throughout the galactic disk, disk-halo interface, and inner CGM (out to 100~kpc). Furthermore, we utilize the star formation and feedback model developed for the IllustrisTNG simulations which is more sophisticated \rev{and widely tested than many of the previous volume-refined CGM-refinement simulations}{than the model used in the \textsc{tempest} and FOGGIE simulations} \citep{pillepich18}. We also include the effects of supermassive black holes and active galactic nuclei (AGN; \citealt{weinberger17}). \rev{}{The simulation files and outputs are made publicly available via Globus\footnote{\url{https://www.scottlucchini.com/engawa}}.}
	
	In Section \ref{sec:methods}, we describe the methodology of the ENGAWA simulations. In Sections~\ref{sec:global}, \ref{sec:mocks}, and \ref{sec:clouds}, we explore the effects of changing resolution on global galaxy properties, on ion column densities, and on the cold cloud population in the CGM, respectively. In these sections, we explore one of our simulations in particular, Auriga halo 6. We then extend the analysis across our four different galaxies in Section~\ref{sec:4gals}. Sections~\ref{sec:discussion} and \ref{sec:conclusions} contain a discussion of our results and our conclusions.
	Throughout this paper, all quantities listed are in physical units unless otherwise specified (e.g. comoving kiloparsecs as ckpc).
	
	\section{The ENGAWA Simulations} \label{sec:methods}
	
	This initial release of the ENGAWA suite consists of cosmological zoom-in simulations with initial conditions (ICs) derived both from the Auriga project \citep{grand17} and from the TNG50 simulation of the IllustrisTNG simulation suite\footnote{\url{https://www.tng-project.org}} \citep{nelson19,nelson19b,pillepich19}. We selected Auriga halos 6 and 8, and TNG50-1 subhalo IDs 537941 and 519311. The Auriga ICs were taken from the public release of the ``Original'' level 4 resolution simulations \citep{grand24}\footnote{\url{https://wwwmpa.mpa-garching.mpg.de/auriga}}. For the TNG50 ICs, all particles within 4 virial radii at $z=0$ were traced backwards to the cosmological ICs at $z=127$, and the Lagrangian region containing all those particles was retained as the high-resolution region (with mass resolution equal to TNG50-1) while the remainder of the cosmological box was degraded.
	
	These zoom-in simulations were run with parameters similar to the TNG50-1 simulation with the exception of kinetic (radio mode) AGN feedback which is not included. We find that the kinetic AGN feedback mode does not alter the galaxy evolution at these mass scales as it mainly plays a role in quenching more massive galaxies.
	Gas cooling and heating is accounted for via atomic and metal-line radiative processes, while additional heating is supplied by a spatially uniform, time-dependent metagalactic UV background tabulated in \citet{uvbg}.
	Black hole seeding, accretion, growth, and feedback is implemented following \citet{weinberger17}. Central supermassive black holes are seeded at a mass of $1.2\times10^6$~\msun\ within galaxies once they reach halo masses of $7.4\times10^{10}$~\msun. They grow according to Bondi accretion (Eddington limited) and emit thermal feedback accordingly based on their accretion rates. Star formation and stellar feedback is based on the \citet{springel03} effective equation of state model and includes a density-dependent pressure floor corresponding to $>10^4$~K. Type Ia and type II supernova feedback is included as well as AGB winds and the corresponding metal enrichment with explicit tracking of H, He, C, N, O, Ne, Mg, Si, and Fe \citep{pillepich18}. Magnetic fields grow from an initial seed field of $10^{-14}$ Gauss comoving through ideal magnetohydrodynamics \citep{pakmor11}. We do not include cosmic rays or thermal conduction. The galaxy properties and default mass resolutions are listed in Table~\ref{tab:galaxies}.
	
	Throughout the first half of this paper (Sections~\ref{sec:global}, \ref{sec:mocks}, and \ref{sec:clouds}) we discuss the properties of Auriga halo 6 (Au6) specifically. We then explore the galaxy$-$galaxy variation across our four different ICs in Section~\ref{sec:4gals}.
	
	\subsection{Volume Refinement Criterion}
	
	All simulations were originally run with traditional mass-based refinement with baryonic mass resolution of $5.4\times10^{4}$~\msun\ (and dark matter mass resolution of $2.9-4.5\times10^{5}$~\msun). Throughout this paper, these simulations are labeled ``Default.''
	Starting from redshift 0.3, we activated an additional, volume-based refinement criterion.
	This starting redshift was chosen to minimize computational cost while still allowing for the galaxy to come to equilibrium at the new resolution. Redshift 0.3 corresponds to 3.4~Gyr ago, which \rev{allows for a sufficient number of sound crossing and dynamical times}{is approximately three sound crossing or dynamical times}. We discuss this further in Section~\ref{sec:discussion}.
	This method refines gas cells surrounding the galaxy based on their volume relative to a specified target gas cell volume. To determine the position of the galaxy, we use the technique developed in \citet{gible} in which the first black hole that is seeded in the simulation is tagged as the central black hole. All radii are then calculated with respect to the position of this black hole particle.
	
	Four additional parameters are introduced for enhancing the spatial resolution around the galaxy $-$ the inner refinement radius, $R_{ri}$, the outer refinement radius, $R_{ro}$, the target gas cell volume, $V_\mathrm{target}$, and the maximum target gas cell volume, $V_\mathrm{target,max}$. For $r<R_{ri}$, cells are refined (derefined) if their volume is more than twice (less than half) $V_\mathrm{target}$. For $R_{ri}<r<R_{ro}$, the target cell volume is interpolated to smoothly transition from the refined region ($V_\mathrm{target}$ at $r=R_{ri}$) out to the rest of the zoom region ($V_\mathrm{target,max}$ at $R_{ro}$). The radially dependent target cell volume, $\tilde{V}(r)$ is determined with the following equation, which assumes a linear transition in cell size. This leads to a smoother transition than a linear relationship in volume:
	\begin{align}
		m &= \frac{(V_\mathrm{target,max})^{1/3}-(V_\mathrm{target})^{1/3}}{R_{ro}-R_{ri}} \\
		b &= (V_\mathrm{target})^{1/3}-m\times R_{ri} \\
		\tilde{V}(r) &= \left(m\times r+b\right)^{3} \, .
	\end{align}
	To be the most computationally efficient, we specify the target volumes in physical units (so that they are larger at earlier times), and the inner and outer radii in comoving units (so that they are smaller at earlier times).
	
	\begin{figure}
		\centering
		\includegraphics[width=0.8\columnwidth]{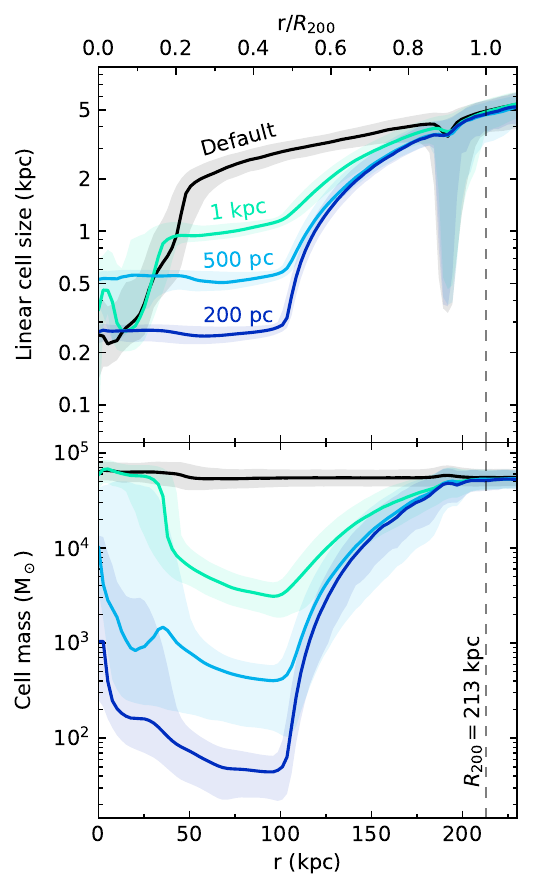}
		\caption{Spatial (top) and mass (bottom) resolution as a function of galactocentric radius at $z=0$. The different colored lines show the median cell size or mass and the shaded regions show the 16\%--84\% range. \rev{}{The vertical dashed line denotes the virial radius of the galaxy.} The \rev{large small size}{} spike at $r\sim200$~kpc corresponds to a satellite galaxy.}
		\label{fig:resolution}
	\end{figure}
	
	\begin{figure}
		\centering
		\includegraphics[width=0.8\columnwidth]{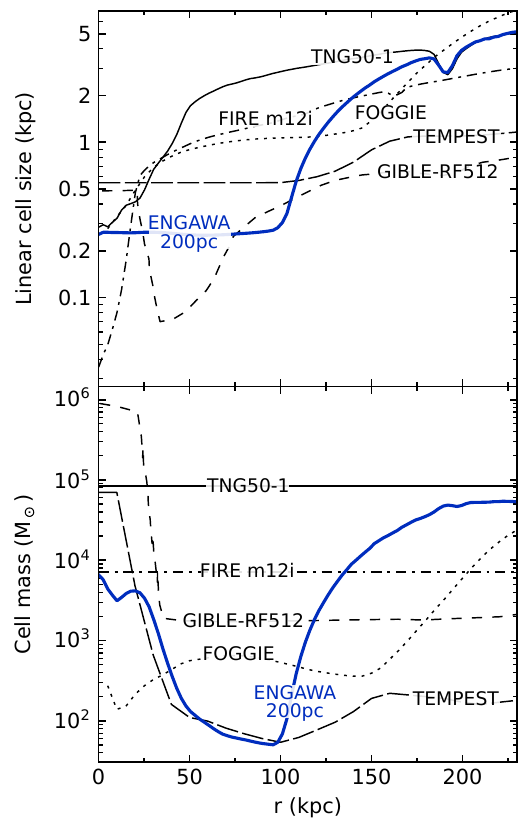}
		\caption{Mean spatial (top) and mass (bottom) resolution as a function of galactocentric radius comparing the 200~pc ENGAWA simulation (red line) against other published simulations at $z=0$. The solid and dot-dashed lines show a cosmological simulation and a standard zoom-in: a galaxy at TNG50-1 resolution, and the m21i galaxy from the FIRE simulations, respectively. The other lines show CGM-refinement simulations: FOGGIE in the dotted line, \textsc{tempest} in the long dashed line, and GIBLE in the dashed line.}
		\label{fig:res-lit}
	\end{figure}
	
	This target volume is then combined with the default target mass criterion to determine whether a gas cell should be refined or derefined. The refinement logic is:
	\begin{lstlisting}
		if (r < R$_\mathrm{ro}$):
		if (vol > 2*targetVol):
		refine()
		if (mass > 2*targetMass):
		refine()
		return
	\end{lstlisting}
	And the derefinement logic is:
	\begin{lstlisting}
		if (r < R$_\mathrm{ro}$):
		if ((vol < 0.5*targetVol) and
		(mass < 0.5*targetMass)):
		derefine()
		else:
		return
		if (mass < 0.5*targetMass):
		derefine()
		return
	\end{lstlisting}
	Here \texttt{r}, \texttt{vol}, and \texttt{mass} are the radius, volume, and mass values for the active gas cell; \texttt{targetVol} is equal to $V_\mathrm{target}$ if $r<R_{ri}$ and $\tilde{V}(r)$ for $r>R_{ri}$; and \texttt{targetMass} is the \texttt{ReferenceGasPartMass} value specified in the parameter file, or calculated at the beginning of the simulation.
	
	\begin{figure*}
		\centering
		\includegraphics[width=1.0\linewidth]{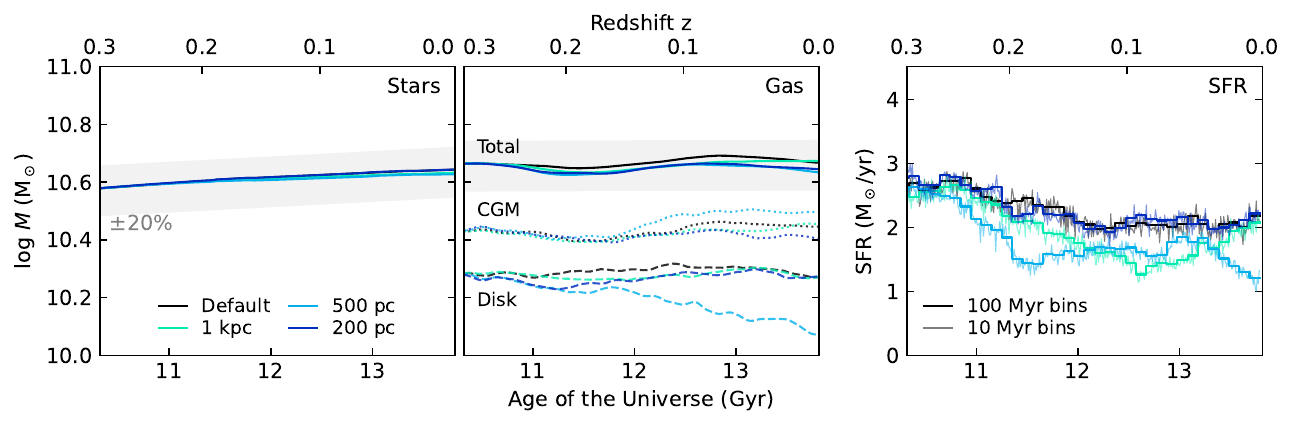}
		\caption{Mass and star formation rate evolution of the Au6 galaxy at four different resolutions. The left panel shows stellar mass, the center panel shows gas mass, and the right panel shows star formation as a function of time from redshift 0.3 when the refinement scheme is activated. In the center panel, the total gas mass is shown by a solid line, and it is also subdivided into disk material (dashed line; gas within $R_{xy}<30$~kpc and $|z|<5$~kpc) and CGM material (dotted line). The gray regions in the left two panels show the $\pm20$\% range following the evolution of the default refinement simulation (blue). In the right panel, the star formation rates are shown binned at 100~Myr intervals for all simulations, with additional thin lines shown for the Default and 200~pc resolution simulations binned at 10~Myrs.}
		\label{fig:mass-comp}
	\end{figure*}
	
	We ran up to three simulations with different target cell volumes in addition to the run with the default mass-based refinement scheme. The radii and maximum target volume values were fixed for the three runs at $V_\mathrm{target,max}=100$~kpc$^{3}$, $R_{ri}=100$~ckpc, and $R_{ro}=200$~ckpc. $V_\mathrm{target}$ was set to 1, 0.125, and 0.015~kpc$^3$ resulting in spatial resolutions of 1~kpc, 500~pc, and 200~pc. The finest resolution was selected such that the average cell sizes within the ISM remain the same between all runs.
	
	Figure~\ref{fig:resolution} shows the linear cell size and cell mass for Au6 as a function of galactocentric distance. The median values are shown as solid lines and the shaded regions show the 16$-$84\% ranges. \rev{}{The top axis shows the radial distance in units of the virial radius, $R_{200}=213$~kpc, and the dip in the linear cell size at $r\sim 200$~kpc is due to the presence of a satellite galaxy.} In the inner regions of the galaxy ($r\lesssim30$~kpc), we see that for the Default and 1~kpc runs the curves are similar due to the fact that the cell sizes in the galactic disk are set by the mass refinement scheme. For the 500~pc run, the median cell size is larger, while the median cell mass remains lower than the default value. This is because these quantities have been calculated with respect to 3D spherical radius and for the 500~pc run, the disk--halo interface and inner CGM contribute an equal or greater proportion of cells as compared with the disk. Therefore, we are not exclusively seeing the disk component at these inner radii. For the 200~pc run, the resolution remains constant throughout the inner 100~kpc of the galaxy, resulting in a smooth transition from the galactic disk out into the inner CGM.
	
	\begin{deluxetable*}{lccccccccc}
		\tablecaption{Present-day galaxy properties of Au6}
		\label{tab:props}
		\tablehead{ & \colhead{$M_*$} & \colhead{$R_{200}$} & \colhead{$M_{200}$} & \colhead{$\dot{M}_*$} & \colhead{$M_\mathrm{gas}$} & \colhead{$M_\mathrm{gas}(\mathrm{H\textsc{i}})$} & \colhead{$M_\mathrm{gas}^\mathrm{CGM}$} & \colhead{$M_\mathrm{gas}^\mathrm{CGM}(\mathrm{H\textsc{i}})$} & \colhead{$N_\mathrm{clouds}$} \\
			& ($10^{10}$~\msun) & (kpc) & ($10^{10}$~\msun) & (\msun~yr$^{-1}$) & ($10^{10}$~\msun) & ($10^{10}$~\msun) & ($10^{10}$~\msun) & ($10^{10}$~\msun) & }
		
		\startdata
		Default & 4.4 & 213 & 102 & 2.1 & 4.7 & 1.6 & 2.8 & 0.03 & 268 \\
		1~kpc & 4.3 & 213 & 103 & 2.1 & 4.7 & 2.3 & 2.8 & 0.12 & 512 \\
		500~pc & 4.2 & 212 & 102 & 1.3 & 4.3 & 1.9 & 2.8 & 0.12 & 2894 \\
		200~pc & 4.4 & 213 & 103 & 2.2 & 4.4 & 2.2 & 2.7 & 0.17 & 18437
		\enddata
		
		\tablecomments{$M_\mathrm{gas}$ is calculated for material within 100~kpc of the galaxy. $M_\mathrm{gas}^\mathrm{CGM}$ is calculated for material between 50 and 150~kpc. $N_\mathrm{clouds}$ is the number of contiguous groups of gas cells with temperatures below $10^{4.5}$~K (see Section~\ref{sec:clouds}). Increase in $M_\mathrm{gas}^\mathrm{CGM}(\mathrm{H\textsc{i}})$ and $N_\mathrm{clouds}$ is due to additional resolved clumping (see Figures~\ref{fig:money} and \ref{fig:smallscale}).}
	\end{deluxetable*}
	
	Figure~\ref{fig:res-lit} shows the mean linear cell size and cell mass of our Au6 200~pc resolution simulation compared against other modern cosmological zoom-in simulations. TNG50-1 is the only full cosmological box simulation shown for reference (as a solid black line). The FIRE-2 m12i galaxy \citep{wetzel16,hopkins18} is shown as a dot-dashed line and while it has a factor of $>10\times$ better mass resolution than TNG50 (bottom panel), the cell size in the CGM is comparable due to its lower densities. The dotted lines show the gas properties for the FOGGIE simulations \citep{foggie} which have fixed spatial resolution of $\sim1$~kpc in a $\pm200$~kpc~$h^{-1}$ box around several MW-mass galaxies. \textsc{tempest} \citep{hummels19} is also shown with long dashed lines reaching $\sim500$~pc resolution. Another simulation specifically designed to resolve the CGM is the GIBLE suite \citep{gible}. These simulations use very high resolution mass refinement in a region from $0.15-1.0\ R_\mathrm{200}$. Although its mean spatial resolution drops below 100~pc, this is dominated by the cold phase. Additionally, within $0.15\ R_\mathrm{200}$ the resolution transitions down to the TNG50-2 level.
	
	Three aspects of our new ENGAWA suite differentiate it from the existing simulations in the literature: (1) We have fixed volume refinement at a higher resolution level \rev{}{at redshift zero} (200~pc) than in any previous work. (2) We maintain this resolution consistently from the disk through the inner CGM to 100~kpc. (3) We use the \rev{sophisticated and }{}well-studied IllustrisTNG feedback model including AGN.
	
	\section{Global Properties} \label{sec:global}
	
	Due to the fact that we have selected our resolution levels so that the ISM remains relatively unchanged, the galaxy itself is similar between different runs. Figure~\ref{fig:mass-comp} shows the time evolution of the galaxy properties. The left panel shows the stellar masses which are all within 4\% of each other at $z=0$. The middle panel shows the evolution of the total gas in the solid lines, then decomposed into the disk material as the dashed lines ($|z|<5$~kpc and $R_{xy}<30$~kpc) and the CGM material as the dotted lines. The right panel shows the star formation rates (SFRs) binned at the level of 100~Myr. \rev{For the default mass-based refinement and the 200~pc refinement simulations, we also show the SFRs binned in 10~Myr as a thin, semi-transparent line.}{The thinner, semi-transparent lines show the SFRs in 10~Myr bins.}
	\rev{}{Additionally,} Table~\ref{tab:props} outlines the properties of the Au6 halo at our different resolution levels. Note that these values differ slightly from those in Table~\ref{tab:galaxies} as \rev{theses}{these} come from the zoom-in simulations run for this work, while the values in Table~\ref{tab:galaxies} are from the original source simulations (TNG50 and Auriga).
	
	The total masses (in stars, dark matter, and gas) are all very consistent. The $z=0$ SFR for the 500~pc simulation is lower than the rest, \rev{however we see variations on the order of a factor of 2 across all resolutions at different times (Figure~\ref{fig:mass-comp}). This also}{which} corresponds with a decrease in the disk gas mass and an increase in the CGM gas mass \rev{which is due to}{because} the gaseous galactic disk \rev{increasing}{increases} in radius. \rev{}{We believe this is due to the additional resolution that is added into the low density regions in the ISM. While the star-forming cells are not changing resolution, the low density bubbles and interarm regions are gaining resolution with the volume refinement active, and this can affect the distribution of gas within the disks and therefore the SFRs.}
	Crucially, we see that the total amount of \hi\ gas is consistent across resolutions (varying by up to 30\%).
	This is notable because the structure and morphology of this material is drastically different depending on the resolution, and the projected column densities in \hi\ increase by $\lesssim4$ orders of magnitude between the highest and lowest resolution simulations.
	
\begin{figure*}
		\centering
		\includegraphics[width=0.8\textwidth]{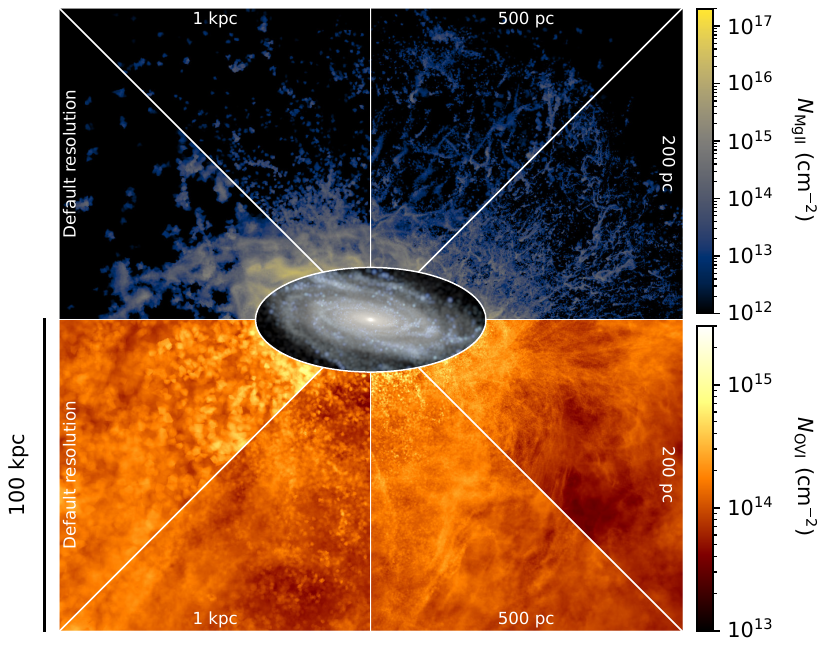}
		\caption{\mgii\ and \ovi\ column densities across different resolutions for the Au6 galaxy at $z=0$ (using \trident\ ion fractions). The central image shows a mock Hubble observation of the stellar component of the galaxy at the same scale. The image is 200~kpc across. The top four panels show the \mgii\ column density for the Default, 1~kpc, 500~pc, and 200~pc resolutions from left to right. The bottom panels show the same for \ovi.}
		\label{fig:money}
	\end{figure*}

	Figure~\ref{fig:money} shows projections of \mgii\ and \ovi\ for the different resolutions of Au6. \rev{}{These column densities are calculated with \trident\ as described in Section~\ref{sec:mocks}.} The central cutout shows a mock \rev{HST}{optical} image of the \rev{}{500~pc run of the} galaxy \rev{}{with the red, green, and blue channels corresponding to the u, g, and r bands, respectively.} \rev{}{The luminosities are }computed based on the ages \rev{}{and metallicities} of the stellar particles \rev{}{assuming a \citet{chabrier03} initial mass function} and tracking extinction from dust, which traces the metals\footnote{\rev{}{This mock optical image was made using the \texttt{image\_maker} tool available at \url{https://bitbucket.org/phopkins/pfh_python}}}. Here we can see that in the cool phase (\mgii), increasing the resolution increases the fragmentation of the structures and results in a higher covering fraction. Interestingly, in the hot phase (\ovi), the material is distributed more smoothly. \rev{There are still well-resolved structures however, and the cell to cell variation in \ovi\ column is lower as the resolution is increased}{At higher resolution, the column density is no longer dominated by a few particularly metal rich cells, and the existence of intricate, filamentary structures appears}. \rev{This seems to be due to the fact that with fixed volume refinement, the hot phase sees a much more dramatic improvement in resolution when compared with the cold phase. Thus, the \ovi\ bearing gas is more well sampled and depends much less strongly on the specific metallicities and temperatures of each individual cell.}{}
	
	\subsection{Small Scales}
	
\begin{figure}
		\centering
		\includegraphics[width=0.8\linewidth]{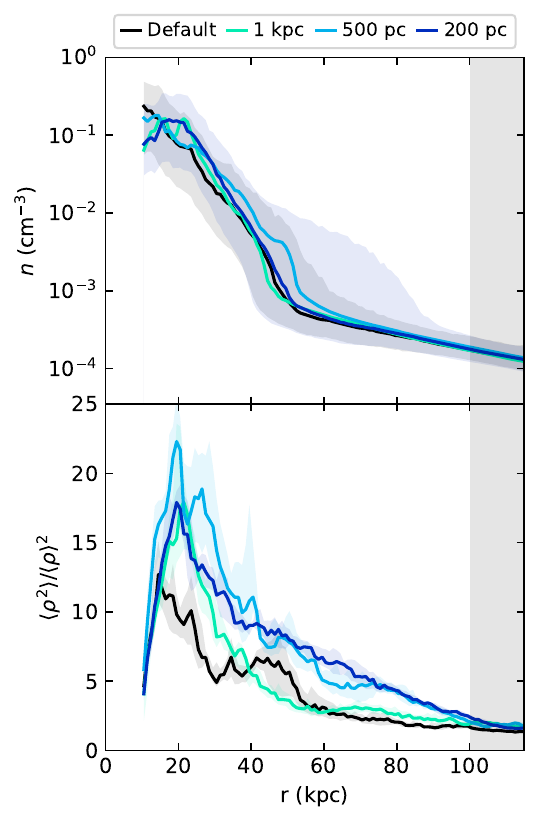}
		\caption{Gas density (top) and clumping factor (bottom) as a function of galactocentric radius at $z=0$. \rev{These are averaged profiles using the last three snapshots in the simulation covering $\sim100$~Myr}{These profiles are averaged over the past 500~Myr, and include only material outside the disk (a cylinder 25~kpc in radius and 10~kpc in height).} In the top panel, each colored line shows the \rev{}{time-averaged median of the} median densities with the shaded regions indicating the \rev{}{time-averaged median values of the} 16--84\% range \rev{}{(for the Default and 200~pc runs)}. The bottom panel shows the \rev{}{time-averaged median} $\langle\rho^2\rangle/\langle\rho\rangle^2$ clumping factor \rev{, increasing with refinement}{with the shaded regions denoting the 16--84\% variation of the clumping factor in time}. The gray shaded regions denote the 100~kpc limit of our full refinement region.}
		\label{fig:smallscale}
	\end{figure}

	We characterize the small-scale properties of the simulated CGM via two different metrics: clumping factor and the second order velocity structure function (VSF). The clumping factor is calculated as $\langle\rho^2\rangle/\langle\rho\rangle^2$, where $\langle\rangle$ is the volume-weighted mean and $\rho$ is the mass density. \rev{The averages are taken over all cells within a given radial bin across the last \rev{3}{11} snapshots of the simulation (covering $\sim100$~Myr) and the results are shown in Figure~\ref{fig:smallscale}.}{For each of the last 11 snapshots of the simulation (covering $\sim500$~Myr), we have computed the median 3D density profile and the clumping factor. In order to isolate the CGM, we have excised a cylinder 25~kpc in radius and 10~kpc in height from these calculations. The solid lines in Figure~\ref{fig:smallscale} show the median radial profile across all the different snapshots while the shaded regions show the 16--84\% range in temporal variation.}
	The top panel shows the 3D radial density distribution, which is very consistent across the different resolution simulations. \rev{}{The 500~pc simulation lies slightly above the other curves due to its disk being slightly more extended.} The bottom panel shows the clumping factor calculated for each of the simulations. While there is significant variability, the higher resolution simulations show generally higher clumping factors.
	
	In the $50<r<100$~kpc range (beyond the galactic disk, and within our enhanced refinement region), we see a monotonic increase in clumping factor with resolution. However, the difference is much less dramatic between the 500~pc and 200~pc resolution simulations than with lower resolutions. Specifically, the clumping factor averaging over this full radial range is \rev{4.2, 6.6, 15.3, and 13.7}{$3.0\pm0.4$, $3.2\pm0.1$, $6.7\pm0.4$, and $5.7\pm0.3$} for the default to highest resolutions, respectively, \rev{}{with uncertainties denoting the 16--84\% range across the 11 snapshots}. While further tests are required, this seems to indicate that this metric for CGM morphology may be converged as we approach resolutions of a few hundred parsecs.
	
	We also use a second order VSF to characterize the strength of resolved turbulence in the simulation as a function of resolution. As defined in \citet{zuhone16}, the VSF is computed as the mean of the squared velocity differences between points at a given spatial separation (in each dimension):
	\begin{equation}
		\mathrm{VSF_2}(r) = \langle|\vec{v}(\vec{\chi}+\vec{r})-\vec{v}(\vec{\chi})|^2\rangle \, .
	\end{equation}
	Here $r=|\vec{r}|$, and $\langle\rangle$ denotes the average over all $\chi$; i.e. the average over all points in the snapshot with a fixed separation $r$.
	For the actual computation of the VSF, we \rev{calculate velocity differences for randomly selected pairs of points across the simulation}{deposit the particles onto a grid using \texttt{yt}'s \texttt{covering\_grid} function. The grid size is set to be the smallest power of 2 (across a 400~kpc box) that is larger than the median cell size in the simulation. This results in grid sizes of 3.1, 1.6, 0.8, 0.4~kpc for the Default, 1~kpc, 500~pc, and 200~pc runs, respectively. We then compute the velocity differences for randomly selected grid cells. Since this is a uniform grid, this results in an effective volume weighting}. We start with 10$^4$ random \rev{gas}{grid} cells (restricted to $40<r<100$~kpc, outside the disk, within the full refinement region), and for each of these cells, we select an additional 10$^4$ gas cells to pair with it. However for this second selection, we weight the particles by the inverse of their squared distance to ensure that we are fully populating the VSF at small separations \citep{fournier25}. This gives us $10^8$ pairs of points which we bin as a function of separation and calculate their average squared velocity differences per bin.
	
\begin{figure}
		\centering
		\includegraphics[width=0.8\linewidth]{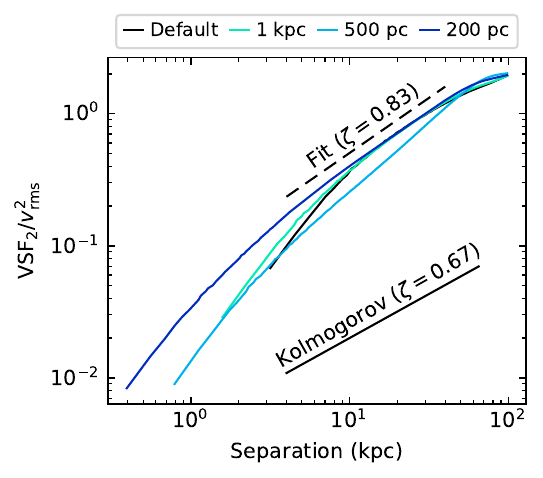}
		\caption{Second order velocity structure function for the different resolution simulations of halo Au6 for the region $40<r<100$~kpc, averaged over the past 350~Myr\rev{}{, normalized by the velocity dispersion}. \rev{Only those separation bins containing $>$100 pairs of cells are shown, as lower cell numbers are noisy.}{} A Kolmogorov slope ($\zeta=2/3$; \citealt{kolmogorov41}) is shown as a black \rev{line.}{ solid line, and the fit slope is shown as a black dashed line ($\zeta=0.83\pm0.02$) with the uncertainty coming from the standard deviation of fit values at different times.} As the resolution increases, we are able to resolve smaller scales and we see increased power at these scales.}
		\label{fig:vsf}
	\end{figure}

	Figure~\ref{fig:vsf} shows the results of these calculations for the different resolutions of our Au6 galaxy \rev{}{averaged over the past 350~Myr}. \rev{The default mass-based refinement simulation is excluded as computing VSFs for data with variable cell sizes can introduce artifacts. We are only showing data points which were calculated with more than 100 pairs.}{Each of the curves are normalized by the square of the velocity dispersion.} \rev{Clearly,}{By construction of the covering grid,} the minimum separations between cells decrease as we increase the resolution. Additionally, we see the low-end turnover of the VSF being pushed to lower and lower separations as well, indicating that there is more small-scale power in the higher resolution simulations.
	
	\rev{Interestingly, the slope of the VSF changes for the different resolution simulations. We see that moving from 1~kpc to 500~pc resolution, the power is increased on all scales, however, moving from 500~pc to 200~pc resolution, we see a steepening, only increasing the power at smaller scales.}{We see excellent agreement between the slopes of the Default, 1~kpc, and 200~pc simulations in the overlapping inertial range ($\sim10-30$~kpc). Fitting the 200~pc simulation curve in this range, we find a slope of $0.83\pm0.02$ (with the uncertainty coming from the standard deviation of the fit results for snapshots at different times). This is slightly steeper than the value from \citet{kolmogorov41} of 2/3, however} a steeper slope may be more consistent with the inclusion of magnetic fields \citep{goldreich95}. \rev{}{Furthermore, our 3D volume-weighted VSF is fundamentally different from the projected 2D luminosity-weighted VSF that is calculated in observations, so further work is required to fully understand the slope that we find here \citep{mohapatra22,fournier25}.} \rev{however since all of our simulations are full MHD (albeit ideal MHD) it is unclear why the slope would be different for the different resolutions.}{For separations smaller than this inertial range, we see that the power drops at increasingly lower values as the resolution is improved, as expected. We note that the 500~pc simulation is slightly anomalous in that its power below $\sim50$~kpc separations is reduced compared to the other simulations. We are still exploring the implications of this discrepancy.} 
	
	\rev{}{To explore its temporal variation, we have calculated the VSF for each of eight snapshots covering the last 350~Myr. The average over this time frame is displayed in Figure~\ref{fig:vsf}. To determine the temporal variation of the VSF, we have computed the spread in VSF$_2$/$v_\mathrm{rms}^2$ values across the eight snapshots (normalized by the mean value) as a function of separation. This fractional variation is less than 10\% for the Default, 1~kpc, and 200~pc simulations, and it is less than 16\% for the 500~pc simulation. This corresponds to an uncertainty in the slope fit of 0.02.}
	\rev{There could be a radial or mass bias in the point selection, but future work will explore this effect in more detail and the role magnetic fields may have on the structure of the CGM.}{}
	
	\section{Ion Column Density Profiles} \label{sec:mocks}
	
	To calculate the column densities of specific ions throughout the CGM, we use the \trident\ code\footnote{\url{https://github.com/trident-project/trident}} \citep{hummels17}, which utilizes the \texttt{yt}\footnote{\url{https://yt-project.org}} simulation analysis module in python \citep{turk11}. \trident\ has compiled pre-computed ion fraction tables using the \textsc{cloudy} photoionization code \citep{ferland13} and populates each gas cell with additional information about various ion masses. We have used the default \citet{hm12} UV background with self-shielding (while this is different than the \citet{uvbg} model used for the Arepo simulation, we expect these differences to be minimal, especially at $z=0$). \rev{}{This set of \textsc{cloudy} tables was calculated by taking 1D ``boxes'' across a grid of densities, temperatures, and metallicities, and irradiating one end of the box with the UV background radiation. The length of the box is equal to the cooling length of the gas parcel. Once the system has come to equilibrium (between collisional ionization, photoionization from the UV background, and recombination), the gas properties at the other end of the box are tabulated for lookup within \trident. Local stellar radiation is not accounted for in these tables, however this is handled with the COLT code in Section~\ref{sec:rt}.}
	
	In this work, we use the ion fraction calculations from \trident\ to generate mock images of \hi, \mgii, and \ovi\ in column density. While true emission maps require calculations of the emission measure for specific lines (which scales as $\rho^2$ instead of $\rho$; \citealt{foggie11}), we use these maps to visualize CGM structures in different phases and to compare against absorption spectroscopy observations measuring the column densities of many different ions throughout the CGM \citep[e.g.,][]{werk14}. These maps are generated using the \texttt{vortrace} code\footnote{\url{https://github.com/gusbeane/vortrace}} which traces rays through the simulation volume while accounting for the intrinsic Voronoi mesh. Thus it does not require any particle smoothing or artificial gridding of the simulation output.
	
	These results are presented here as a comparison against the existing literature. We provide more realistic predictions based on post-processing with Monte Carlo radiative transfer in Section~\ref{sec:rt} below.
	While inclusion of radiative transfer only minimally changes the \mgii\ and \ovi\ results, we do see significant changes to the \hi\ column density profiles
	with the inclusion of ionizing radiation from stellar sources in addition to the UV background radiation.
	
\begin{figure*}
		\centering
		\includegraphics[width=1.0\textwidth]{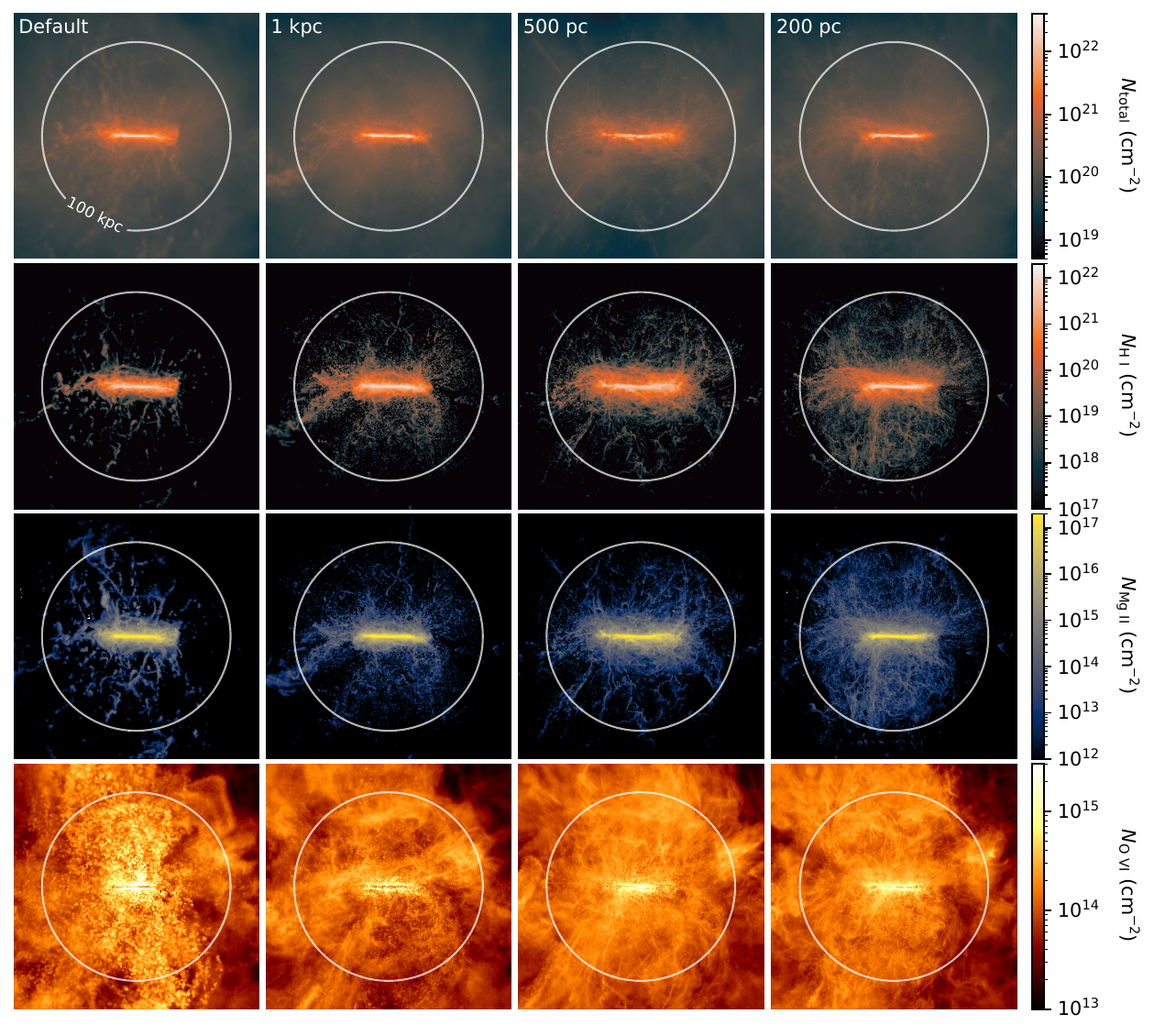}
		\caption{Ion column density projections for four different resolutions of Au6 at $z=0$. From top to bottom, we show total gas, \hi, \mgii, and \ovi\ column densities. From left to right, we show the four different resolution simulations: Default, 1~kpc, 500~pc, and 200~pc. Each image is 260~kpc across with the white circle denoting $r=100$~kpc, the edge of the full refinement region.}
		\label{fig:proj}
	\end{figure*}

	Figure~\ref{fig:proj} shows the edge-on cartesian projections of total gas, \hi, \mgii, and \ovi, from top to bottom, for the four different resolutions of the Au6 galaxy: Default, 1~kpc, 500~pc, and 200~pc, from left to right. The circle in each panel denotes a radius of 100~kpc which is the edge of our full refinement region. As discussed above (see Figure~\ref{fig:money}), we see dramatic increases in column density and structure for \hi\ and \mgii\ as the resolution is improved, while in \ovi, the changes between fixed resolution runs (right three panels) is less dramatic than the change from default mass refinement to 1~kpc volume refinement. Again, this is due to the fact that \ovi\ traces the hot, low-density phase and the cell sizes for this material in the Default run are $\gtrsim2$~kpc. Furthermore, the \ovi\ distribution becomes quite filamentary as the resolution is increased. These filaments may be tracing the edges of large superheated bubbles originating from the high-metallicity material being deposited into the inner CGM from stellar feedback.
	
\begin{figure*}
		\centering
		\includegraphics[width=1.0\linewidth]{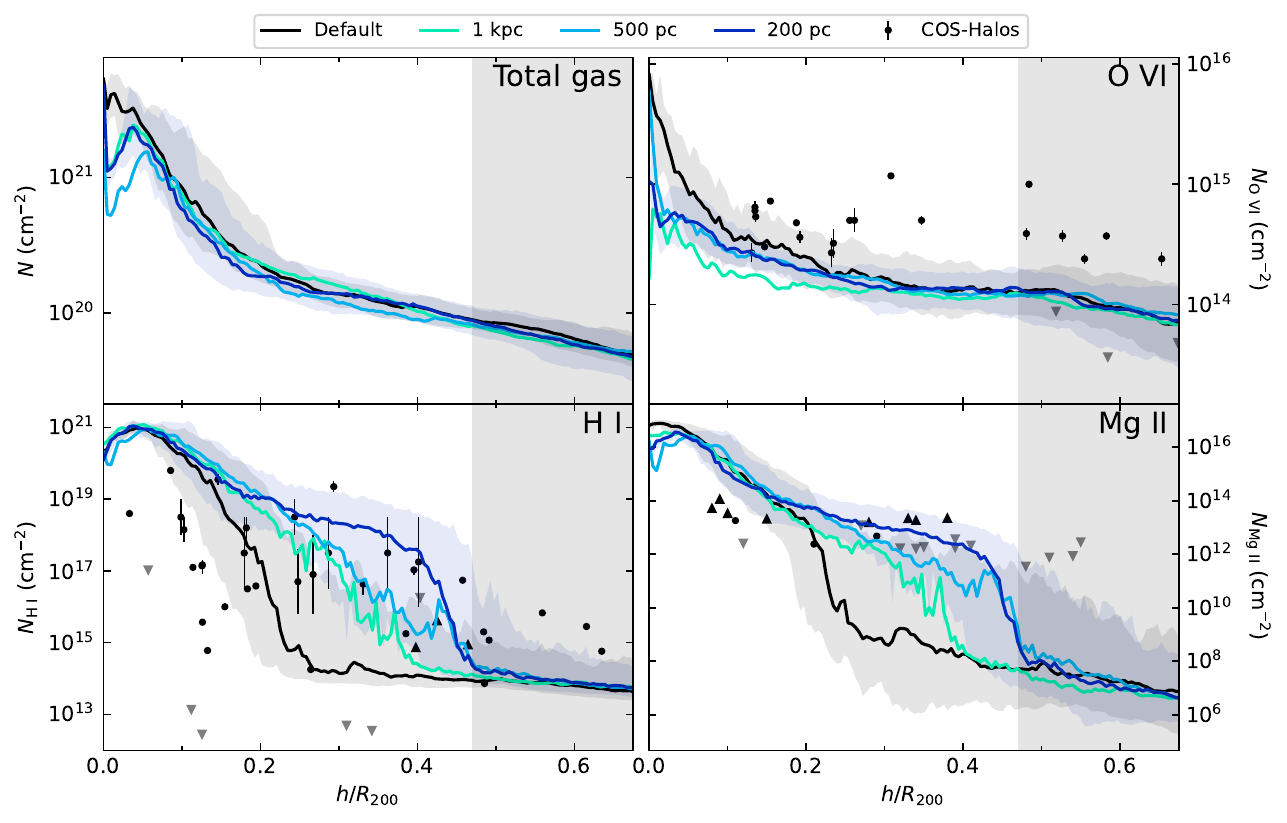}
		\caption{Radial column density profiles for four ions at $z=0$ at an inclination of $\sim20^\circ$. In each panel, the solid lines denote the median values for each simulation and the shaded regions (shown for the Default and 200~pc resolution runs) indicate the 16\%--84\% ranges taken from 360 sightlines at each radius. In the top row, we show the total gas and \ovi\ column densities as a function of impact parameter in units of the virial radius ($R_\mathrm{200}=213$~kpc for Au6), which are relatively unchanged across resolutions. The middle panels show \hi\ and \mgii\, which exhibit a dramatic increase as the resolution is improved. The observed column densities of \hi, \mgii, and \ovi\ from the COS-Halos survey are also included in their respective panels (\hi: \citealt{tumlinson13,prochaska17}; \mgii: \citealt{faerman23}; \ovi: \citealt{tumlinson11}). Upward facing arrows are lower limits, downward facing arrows are upper limits, and circles are measurements with uncertainties. The gray shaded region denotes that the resolution begins to degrade for $r>100$~kpc.}
		\label{fig:coshalos}
	\end{figure*}

	We then quantify the column density profiles of these different simulations in Figure~\ref{fig:coshalos} (for Au6). To produce these figures, we again use \texttt{vortrace} to compute the column densities of 360 sightlines at each projected radius around the galaxies in an inclined orientation ($i\sim20^\circ$). Each of the panels in this figure depict column density as a function of impact parameter with the colored lines showing the median value of the 360 sightlines and the shaded regions showing the 16\%$-$84\% range (only shown for the Default and 200~pc runs for clarity).
	The \hi, \mgii, and \ovi\ panels also include observational data from the COS-Halos survey (\hi: \citealt{tumlinson13,prochaska17}; \mgii: \citealt{faerman23}; \ovi: \citealt{tumlinson11}, note that for the \ovi\ we have only used the star-forming sample). The dots are measurements, while the triangles are upper or lower limits.
	Each panel includes a gray shaded region covering radii beyond 100~kpc ($h/R_{200}=0.47$ for Au6) since that is the limit of our full refinement region.
	
	Figure~\ref{fig:coshalos} quantitatively shows that we see strong enhancements in cold and cool column densities throughout the enhanced refinement region. However, total gas column and \ovi\ column are not much affected. Furthermore, as stated in Section~\ref{sec:global} and Table~\ref{tab:props}, the total \hi\ gas mass in the CGM is only increased by a factor of a few. This is discussed further in Section~\ref{sec:discussion}.

	Importantly, in the region $50<h<100$~kpc, the enhanced refinement simulations are, in general, more consistent with the COS-Halos results for \hi. The 200~pc resolution simulation does over-predict \hi, however this is mitigated upon the inclusion of stellar radiation (see Section~\ref{sec:rt}). This is quantified in Section~\ref{sec:discussion}.
	We note that most of the COS-Halos galaxies are slightly lower mass than fiducial MW values (log~M$_*=10.6$ for Au6 vs. $9.5-11.1$ for the star forming galaxies in the observations), and thus we are comparing against impact parameter normalized by virial radius. It is also worth noting that the simulation is not converged. We see dramatic increases in column density changing from 500~pc to 200~pc resolution. Previous efforts have also found unconverged behavior \citep{gible,hummels19,vandevoort19}, and high resolution simulations and analytic arguments suggest clouds could continue to fragment down to sub-pc scales \citep{mccourt2018}. Conversely, additional physics may be at play to set the sizes of these small-scale structures, such as cosmic rays or galactic ionization \citep{roy25}. In the next section, we do see that the inclusion of stellar radiation tempers this slightly, however we will also continue to explore this in future work.
	
	\begin{figure}
		\centering
		\includegraphics[width=0.43\textwidth]{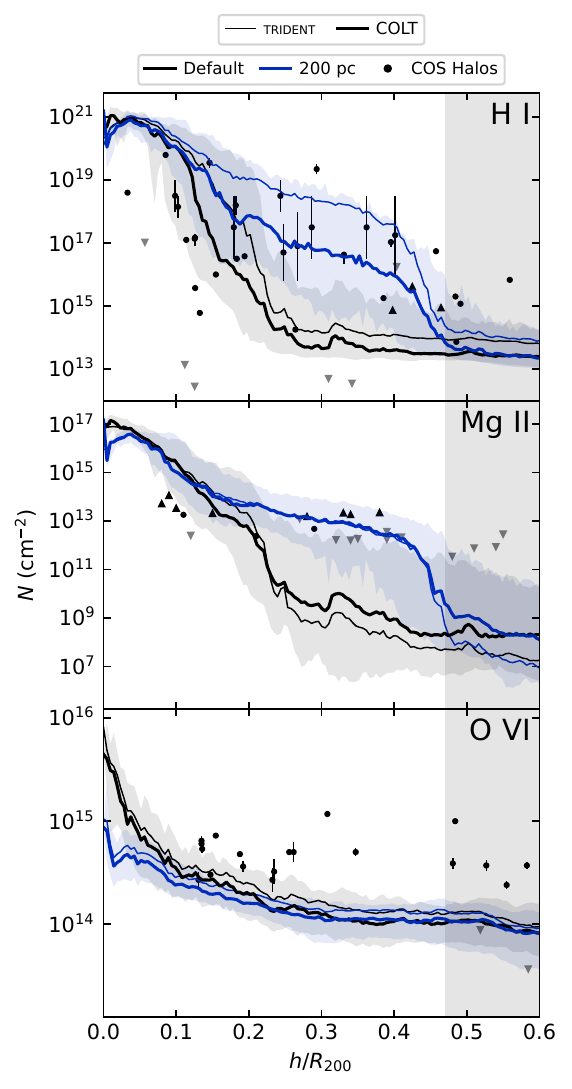}
		\caption{As Figure~\ref{fig:coshalos} comparing \rev{values with and without post-processing radiative transfer}{output from \trident\ (UVB + self-shielding) with COLT (stellar radiation + tabulated UVB)}. The black lines show the Default resolution simulation, while the blue lines show the 200~pc resolution run. The thin lines show the column densities \rev{straight out of the simulation}{computed with \trident} and the thick lines are recomputed after COLT \rev{}{Monte Carlo radiative transfer} post-processing. The shaded region surrounding the lines denote the 16$-$84\% ranges for column densities found at that given radius (for the COLT \rev{postprocessed }{}results). The black points (arrows) are data (limits) from COS-Halos. From top to bottom, the panels show \hi, \mgii, and \ovi\ profiles. The grey shaded region on the right denotes the limit of our full high-resolution region in the galaxy (100~kpc).}
		\label{fig:colt_columns}
	\end{figure}
	
	\begin{figure}
		\centering
		\includegraphics[width=0.43\textwidth]{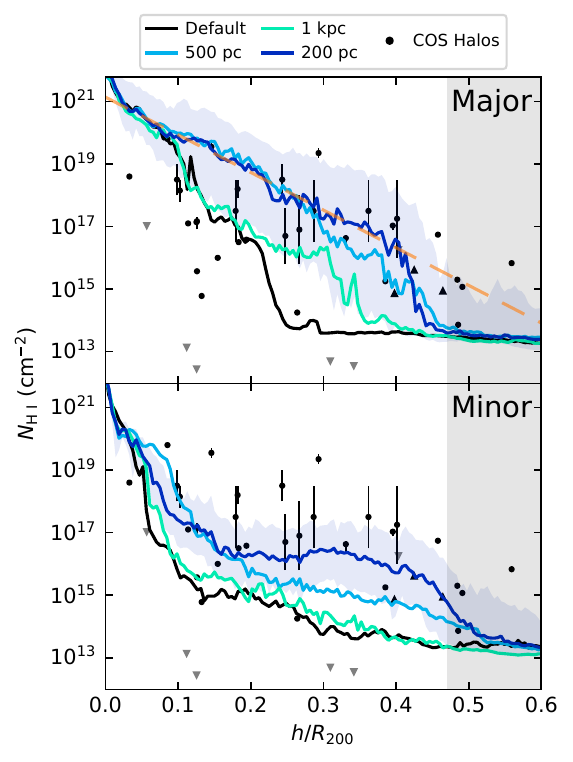}
		\caption{As Figures~\ref{fig:coshalos} and \ref{fig:colt_columns} except showing the COLT\rev{ postprocessed}{} column densities split by major and minor axis for all resolutions. The galaxies were rotated fully edge on for these calculations. The orange dashed line in the top panel is a fit to the 200~pc resolution profile for the range $0.1<h/R_\mathrm{200}<0.4$. The equation for the fit is $\log(N_\mathrm{H\ I})/\mathrm{cm}^{-2}=21.0-11.5\times h/R_\mathrm{200}$.}
		\label{fig:colt_columns_major_minor}
	\end{figure}
	
	\subsection{\rev{Post-processing Radiative Transfer}{Stellar Radiation}} \label{sec:rt}
	
	To more accurately model the abundances on ions around these galaxies, we have used the COLT (COsmic Lyman-alpha Transfer; \citealt{colt, Smith2022}) radiative transfer (RT) code\footnote{\url{https://colt.readthedocs.io}} to include ionizing radiation from stellar sources\rev{ and a UV background}{}. COLT employs a Monte Carlo radiative transfer (MCRT) technique to sample the processes of photon emission, propagation, absorption, and scattering through snapshots from hydrodynamical simulations.
	In this work, we utilize the photoionization equilibrium capabilities of the code to calculate updated ionic abundances for each gas cell. Note that this is a post-processing technique in that these ion fractions are not computed self-consistently as the simulation is evolving, they are adjusted after the fact on a single snapshot. However, due to the increased computational expense of full radiative transfer simulations (e.g., \citealt{thesan,cadiou25,Kannan2025}) we believe this is a good middle ground to provide more realistic results, with a complementary perspective to the results presented above.
	
	\rev{}{While \trident\ included the effect of uniform UV background radiation on the ion fractions of a given gas cell, COLT self-consistently tracks photons as they propagate through the simulation volume. Both tools calculate ion fraction equilibrium given collisional ionization, photoionization (from various sources), and recombination. In COLT, photons are sourced from star particles and modeled via spectral energy distribution functions (SEDs). As the photons pass through gaseous material, their SEDs are modified, and the properties of the gas cells are updated as a result of absorption and emission. This automatically handles self-shielding, however the implementation of the UV background is reduced to the uniform injection of tabulated values for hydrogen and helium (from \citealt{hm12}) with the \citet{Rahmati2013} self-shielding prescription. This is discussed further in Section~\ref{sec:discussion}.}
	
	We supply COLT with the gas and stellar particle information from the simulation including positions, gas densities, internal energies, and metallicities, and stellar initial masses, ages, and metallicities. We assume a dust-to-metal ratio of 0.3 with the MW dust model from \citet{weingartner01} and stellar SEDs following a Chabrier IMF ($\alpha = -2$; maximum mass 100~\msun) including binary stars \citep{stanway18}. Due to the unresolved nature of the ISM in these simulations, we enforce a temperature of $10^4$~K in gas cells with nonzero star formation rates and apply a stellar birth cloud escape fraction of 25\% to mimic the absorption of photons from the unresolved multiphase ISM. The specific choice of 25\% for this parameter ($f_\mathrm{esc,*}$) is discussed in Appendix~\ref{appendix:fesc}.
	
	\begin{figure*}
		\centering
		\includegraphics[width=1.0\textwidth]{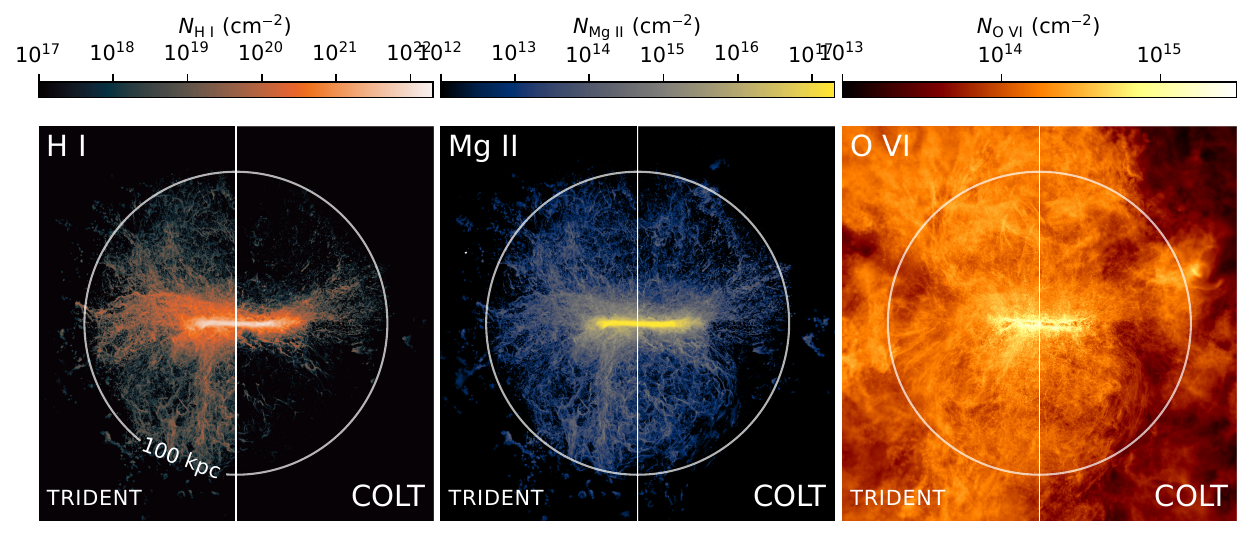}
		\caption{As Figure~\ref{fig:proj} except the left half of each panel shows the 200~pc simulation predicted column densities using \trident, while the right half shows the values when \rev{applying the}{including stellar radiation with} COLT\rev{ post-processing radiative transfer}{}. The circle denotes 100~kpc in radius, the edge of our fully refined region.}
		\label{fig:colt_projection}
	\end{figure*}
	
	With these settings, we achieve ionizing escape fractions ($>13.6$~eV) at the virial radius of $\sim5\%$, in line with estimates for our own Galaxy \citep{dove00}. Reapplying our analysis with \texttt{vortrace}, we see a shift in the predicted column densities for the ions that we have considered. Figure~\ref{fig:colt_columns} shows the column density profiles for \hi, \mgii, and \ovi\ \rev{with (thick) and without (thin) COLT post-processing}{from COLT (thick) and \trident\ (thin)} for the Default (black) and 200~pc (blue) resolution runs. As in Figure~\ref{fig:coshalos}, the galaxy is taken at an inclination of $\sim20^\circ$ and the COS-Halos data points are overlaid \citep{tumlinson11,tumlinson13,prochaska17,faerman23}. The \hi\ columns are reduced due to the increased radiation, however values are actually more consistent with the COS-Halos observations (see discussion in Section~\ref{sec:discussion}).
	For \mgii\ and \ovi, we do not see significant changes. We note that this may be due to the fact that in COLT, the UVB radiation only affects hydrogen and helium, while in \trident, the \textsc{cloudy} runs include the UVB radiation and update all ion fractions accordingly. Thus, the coincidence that the \mgii\ profile for the 200~pc resolution run is the same \rev{with and without RT}{from \trident\ and COLT} is due to the fact that the ionizing energy is supplied from the UVB in \trident\ while it is supplied by the stellar radiation with COLT. \rev{And this happens to be balanced with $f_\mathrm{esc,*}=0.25$. We see different behavior for different values of $f_\mathrm{esc,*}$ as we discuss in Appendix~\ref{appendix:fesc}. So}{For $f_\mathrm{esc,*}=0.25$, these two sources of radiation appear to have the same intensity. Upon inclusion of the effects of a UVB on \mgii\ in COLT,} we would expect the amount of \mgii\ to decrease slightly. \rev{upon the inclusion of both of these sources of radiation.}{We believe the lack of a radial dependence in the difference between the COLT and \trident\ profiles is due to the fact that the \mgii\ profile is largely determined by the Mg density and the fraction of Mg in \mgii\ is determined by the overall strength of the radiation. This can be seen in Appendix~\ref{appendix:fesc} where we see similar profiles for different values of $f_\mathrm{esc,*}$, just differing in normalization.}
	\rev{this \rev{does not seem to be}{is not} the case for \rev{the}{}\ovi}{Due to its high ionization potential, the stellar radiation field does not seem to have a significant effect on the \ovi\ column densities,} as we do not see a decrease in column as the level of radiation is increased (see Appendix~\ref{appendix:fesc}).
	
	Figure~\ref{fig:colt_columns_major_minor} shows the \hi\ column density data for all resolutions split by major and minor axis of the galaxies (for Au6). The galaxies have been rotated fully edge on for these calculations, and major axis contributions are computed from the sightlines with either $-45^\circ<\phi<45^\circ$ or $135^\circ<\phi<225^\circ$. The remaining sightlines ($45^\circ<\phi<135^\circ$ or $225^\circ<\phi<315^\circ$) are used in the minor axis calculation. As in Figures~\ref{fig:coshalos} and \ref{fig:colt_columns}, the solid lines denote the median values and the blue shaded regions show the 16$-$84\% range of column density values for the 200~pc resolution simulation with the black data points show the observational measurements \citep{tumlinson13,prochaska17}. By splitting the simulations up in this way, we can see that the high column density observations are reasonable assuming they are along the major axis of their galaxies, while the lower $N$ data (specifically for $h/R_\mathrm{200}<0.2$) are within the ranges found along the minor axes in the simulations.
	
	Along the major axis of the galaxy, we also see that the column densities appear converged between the 500~pc and 200~pc resolution simulations. Furthermore, there is a nice \rev{linear}{exponential} behavior (out to the point at which the resolution begins to degrade). We have fit the profile of the 200~pc resolution simulation (in the range $0.1<h/R_\mathrm{200}<0.4$) and we find an equation of $\log(N_\mathrm{H\ I})/\mathrm{cm}^{-2}=21.0-11.5\times h/R_\mathrm{200}$.
	
	\begin{deluxetable*}{lcccccc}
		\tablecaption{Resolution Dependence of Cloud Counts}
		\label{tab:clouds}
		\tablehead{ \colhead{Resolution} & \multicolumn{3}{c}{Temperature Cut} & \multicolumn{3}{c}{Density Cut} \\
			& \colhead{$N$} & \colhead{$N^\mathrm{CGM}$} & \colhead{log $M^\mathrm{CGM}$} & \colhead{$N$} & \colhead{$N^\mathrm{CGM}$} & \colhead{log $M^\mathrm{CGM}$}}
		\startdata
		Default & 268 & 130 & 8.73 & 1119 & 315 & 8.65 \\
		1 kpc & 512 & 277 & 8.87 & 2880 & 1526 & 8.84 \\
		500 pc & 2894 & 2371 & 8.91 & 13410 & 9752 & 9.00 \\
		200 pc & 18437 & 14112 & 8.98 & 95747 & 71458 & 9.01
		\enddata
		\tablecomments{CGM superscript indicates the region $50<r<100$~kpc.}
	\end{deluxetable*}
	
	We do note that this azimuthal dependence of the \hi\ columns is not found without COLT post-processing. This is visible in Figure~\ref{fig:colt_projection} which 
	shows edge on projections of column densities \rev{with (right halves of each panel) and without (left halves of each panel) COLT RT}{computed with \trident\ (left halves of each panel) and with COLT (right halves of each panel)}. While the azimuthal variation of \mgii\ and \ovi\ is not dramatically affected, we see a strong reduction of the \hi\ columns along the minor axis of the galaxy. This is expected due to the higher escape fractions along these directions. This could be one of the reasons that we see high-velocity clouds in simulations out at much larger distances into the CGM than in simulations \citep{lucchini24}. \rev{We}{Future work} will explore whether the combination of enhanced resolution and stellar radiation could provide a better fit to the distance distribution of high-velocity clouds.
	
	\rev{Since these calculations are purely determining photoionization equilibrium, the cell temperatures are not updated throughout this process. Future work will also explore thermal equilibrium \citep{mcclymont25},}{While COLT has a full thermal equilibrium solver \citep{mcclymont25}, we have restricted our calculations to fix the gas cell temperatures for this work. Future work will explore this additional dimension,}
	however it is unclear how it will interact with the adopted effective equation of state model. The least disruptive strategy would be to retain the constant temperature requirement for the ISM so only non-star-forming gas is affected. That being said, since the temperatures (and total gas densities) remain fixed, the results in the following section exploring the properties of cold\rev{}{/discrete} clouds apply regardless of whether photoionization equilibrium RT is included.
	
	\section{\rev{Cold}{Discrete} Clouds} \label{sec:clouds}
	
	Clouds are an important feature of the multiphase CGM, and their survival in the hotter, diffuse CGM is still poorly understood. Understanding their interaction with turbulence and thermal winds, gas mixing, and properties is key in determining their role in galactic feedback and star formation \citep{ramesh26}. Higher resolution simulations allow us to better constrain the state of gas mixing around clouds to gain insight on their origin and journey towards the galactic disk.
	
	Following \citet{nelson20} and \citet{lucchini24}, we identify and isolate contiguous cool clouds throughout the CGM to see how their distribution and properties change with varying resolution. To select these clouds, we rebuild the Voronoi mesh from the snapshot output of the simulation at $z=0$ and determine the direct neighbors for each cell. We then look for connected components after masking to only include cells with $T<10^{4.5}$~K. To ensure that we are resolving the structures, we further restrict our definition of a cool cloud to those connected groups with at least 10 cells.
	
	In the original \textsc{tempest} simulation \citep{hummels19} as well as a recent paper from the FOGGIE collaboration \citep{foggie10}, they also explored the cloud distribution selecting based on gas densities. We performed a similar analysis to compare across different hydrodynamics schemes (grid-based in ENZO, Voronoi mesh in Arepo). Again we rebuilt the Voronoi mesh and determined neighbors for each cell, however we then found the connected components after applying masks based on the densities of each cell. We first selected all cells with densities above 10 times the minimum density in the simulation, then increased the density threshold by factors of two until we reached the maximum density (so no more cells were selected). We then constructed a tree in which each connected component at a given density threshold has children that consist of connected components at the next highest density threshold with overlapping gas cells. Finally, we define a cloud as any connected component (at any density threshold) not having any children.
	
	\begin{figure*}[t!]
		\centering
		\includegraphics[width=1.0\linewidth]{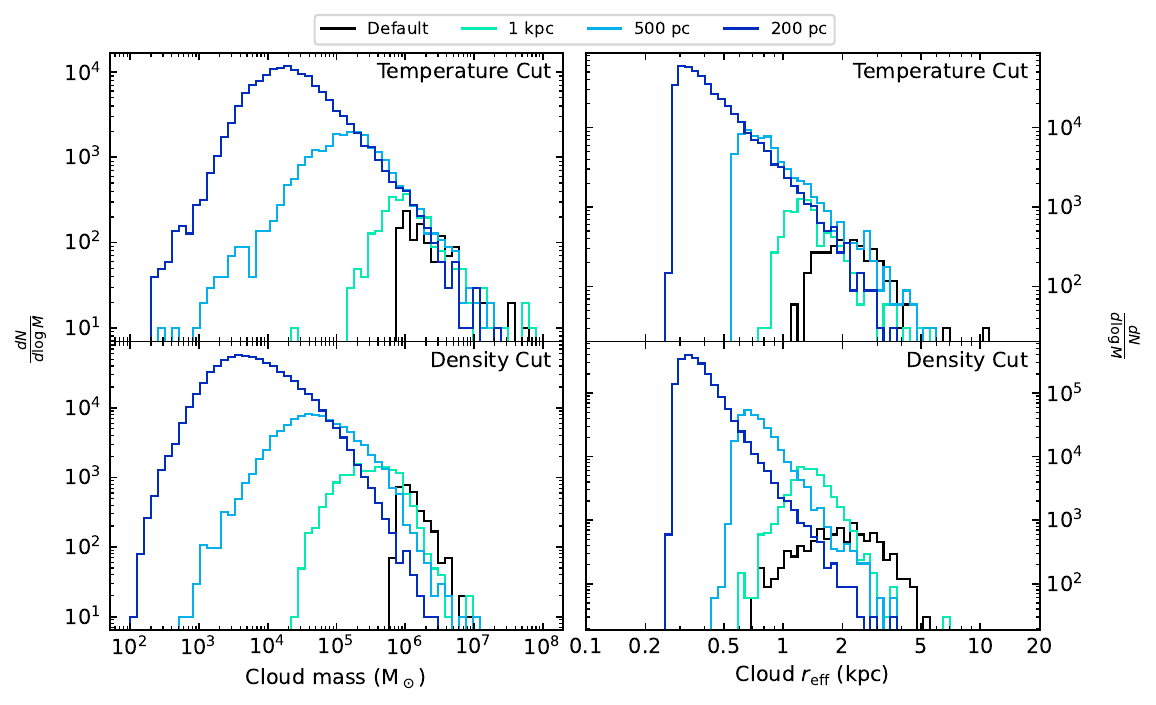}
		\caption{Cloud mass and size compared across identification methods and across resolution. These distributions only show clouds $50<r<100$~kpc $-$ outside the disk, within the full refinement region. Each panel depicts a different cloud identification technique and the colored lines show the distributions for different simulation resolutions of Au6.
			The top panels show the distributions for clouds identified by a uniform temperature threshold. The lower panels show the results for clouds identified based on their relative density to their immediate surroundings. See text for descriptions of identification methods.}
		\label{fig:distributions}
	\end{figure*}
	
	This results in two different populations of clouds each with different properties: contiguous cells with $T<10^{4.5}$~K and clouds defined by local over-densities. With these two different methods for cloud identification, we are able to address the discrepancies highlighted by prior work relying on only one of these two methodologies. The number of clouds identified using each method is outlined in Table~\ref{tab:clouds}, and their mass and size distributions are shown in Figure~\ref{fig:distributions}. The cloud mass is calculated by summing the masses of each individual gas cell included in each cloud, and the effective radius is the radius corresponding to a sphere with the same volume as the cloud: $r_{\text{eff}} = \sqrt[3]{\frac{3}{4 \pi} V_{\text{cloud}}}$. Increasing resolution results in an increase in cloud number in both cases, however there are more over-dense regions defined as clouds across all resolutions. The average cloud mass and effective radius both decrease with increasing resolution, as smaller clouds can be resolved, however Figure~\ref{fig:distributions} illustrates that this change is not uniform across the two methods with resolution.
	
	Using the temperature cut method, we see that the number of massive clouds is consistent across resolutions. For clouds more massive than $\sim10^6$~M$_\odot$, and sizes larger than $1-2$~kpc, the curves in Figure~\ref{fig:distributions}, top panels, are aligned. This is an important result because it indicates that as resolution is increased, the large clouds are \textit{not} fragmenting into many smaller clouds: instead, we are simply able to isolate more smaller clouds.
	However, with the density cut method we do see a decrease in the number of large clouds as a function of resolution (Figure~\ref{fig:distributions}, bottom panels).
	\rev{So clearly, there is some restructuring of the CGM that is happening as the resolution is increased. The largest cool structures remain, however the largest overdense structures fragment. This could simply be a result of the fact that the density cut clouds are determined relative to their surroundings so they are naturally resolution dependent (the simulation can resolve smaller cores at higher resolution), whereas the temperature cut clouds are defined based on a fixed cutoff ($10^{4.5}$~K).}{This is due to the fact that, with the density cut method, a single large cloud could simply gain substructure as resolution is increased. This would remove the large cloud from the catalog (since it is no longer a leaf in the tree) and replace it with a smaller cloud. These results are consistent with previous findings \citep{cadiou25}.}
	
	\rev{}{In fitting the data from the temperature cut cloud mass function, we find a consistent power law slope across all resolutions. We fit the resolved portions of each curve (defined as masses larger than the peak of the histogram) with $dN/d\log M \propto M^{\Gamma}$. Combining all four simulations, we find $\Gamma=-0.95\pm0.13$. This corresponds to a profile of $dN/dM\propto M^{\alpha}$ where $\Gamma=1+\alpha$, such that $\alpha\approx-2$, slightly steeper than slopes of $-3/2$ found for MW high-velocity clouds \citep{wakker91}, or $-1.8$ from \citet{cadiou25} at high redshift, however this discrepancy could be due to the selection methods.}
	
	Table~\ref{tab:clouds} also shows the number of clouds limited to the CGM region $50<r<100$~kpc (columns 3 and 6). This includes 30\% and 50\% of the clouds for the temperature and density cut clouds, respectively, at the default resolution. But increases to include 75\% of the clouds for both populations in the 200~pc resolution simulation. We also sum up the total mass in all the clouds in the CGM region for each of the simulations and both cloud selection methods (columns 4 and 7 of Table~\ref{tab:clouds}). Even across all eight values, we see $\lesssim50$\% variation, but comparing the masses of the temperature cut and density cut clouds, we see agreement ranging from 5\% to 19\%. Therefore, while the total number of clouds between the two identification methods are very different, the total mass is very similar indicating that the larger temperature cut clouds are likely composed of several smaller density cut clouds. \rev{}{This is consistent with the picture that while the clouds are not fragmenting, they are gaining substructure.}
	
	\begin{figure}
		\centering
		\includegraphics[width=0.45\textwidth]{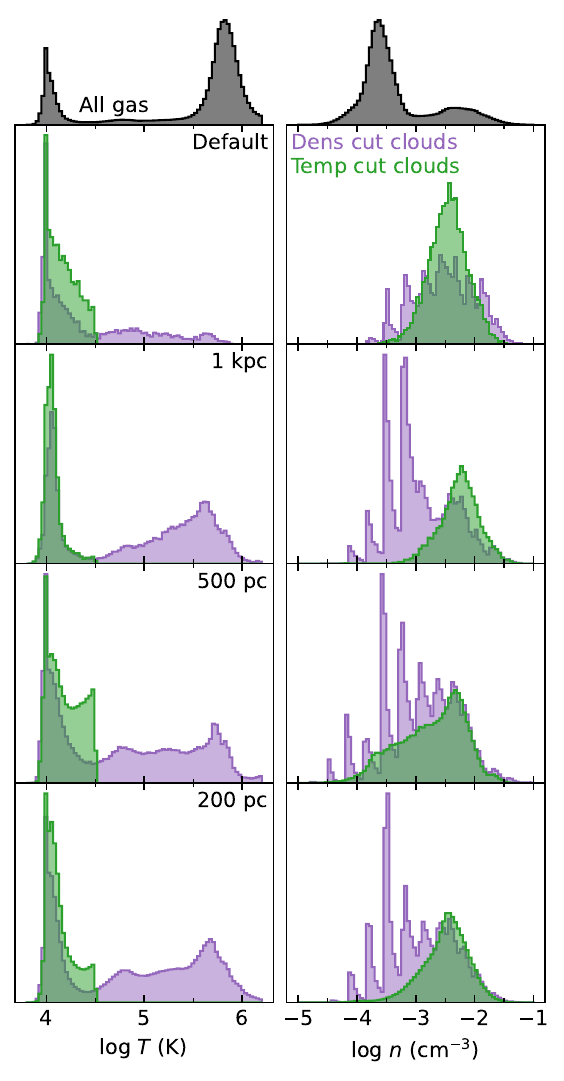}
		\caption{Cloud temperature and density compared across resolution and identification methods. As in Figure~\ref{fig:distributions}, only those clouds within $50<r<100$~kpc are included. In contrast with Figure~\ref{fig:distributions}, each panel depicts a different simulation resolution and the histograms show the property distributions for the density (purple) and temperature (green) cut clouds. Above the four panels, we show the distribution of all gas cells within $50<r<100$~kpc on an arbitrary scale in gray. The left panels show the mean cloud temperatures and the right panels show the mean cloud densities. Here we have included all the individual gas cells that comprise the clouds in the histogram to better account for the difference in the number of clouds identified via the two methods. \rev{}{The spikes in the density distribution for the density-cut clouds are a result of the powers of two separation in density thresholds.}}
		\label{fig:cloudrhot}
	\end{figure}
	
	\begin{figure}
		\centering
		\includegraphics[width=0.8\columnwidth]{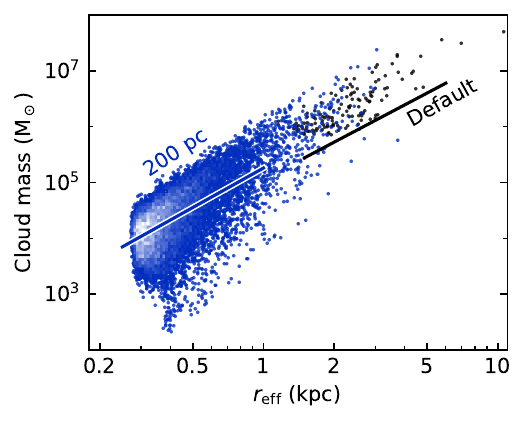}
		\caption{The mass$-$size relation for temperature cut identified clouds. \rev{Contours represent equally spaced levels in the kernel density estimation of the distribution of clouds for each of the resolutions. Linear fits are also shown for the Default and 200~pc resolution simulations in black and blue, respectively.}{Data points (and histogram for high density regions) are shown for the Default (black) and 200~pc (blue) simulations with linear fits to the data shown.} The fit to the default resolution data takes the minimum detectable mass (due to the resolution) into account via censored data (see text). The slopes are $2.22\pm0.23$ and $2.34\pm0.02$ for the Default and 200~pc runs, respectively, and they are consistent within $\lesssim1\sigma$.}
		\label{fig:mass-size}
	\end{figure}
	
	In the GIBLE simulations, which use increasingly high resolution mass-refinement \citep{gible}, the cloud mass distribution resembles an exponential that extends to lower masses and is truncated at the resolution limit. Interestingly, in the FOGGIE simulations that use volume refinement \citep{foggie10}, the cloud mass distribution is more gaussian and the size distribution follows a truncated exponential. We reproduce these results here, showing gaussian distributions for the cloud masses (more obviously for the density-cut clouds). 
	This indicates that these results are robust against numerical hydrodynamics solver (ENZO vs Arepo) and originate from the difference in cloud selection method.
	
	Figure~\ref{fig:cloudrhot} shows the temperature and density distributions of the CGM clouds identified via the temperature cut (green) and the density cut (purple). Clearly these two techniques are isolating different populations of gas. In the density cut selection, there is a significant number of warm/hot clouds ($10^{4.5}<T<10^6$~K) which are explicitly excluded in the temperature cut sample. Correspondingly, the density cut clouds cover a larger range in density since the only requirement is that they have no child clumps of higher density, while the temperature cut selection isolates the denser regions. \rev{}{This also becomes more prominent with improved spatial resolution due to the fact that the low density (and high temperature) regions become highly resolved as well, i.e., we are resolving much more substructure in the diffuse phase. The spikes visible in the density distributions for the density cut clouds are a result of the powers of two used for the tree building.} Interestingly, the cool density cut clouds do not line up exactly with the temperature cut clouds, indicating that selecting purely based on temperature is likely ignoring substructures at a certain scale. \rev{}{We also note that most of the temperature cut clouds are in the self-shielded regime ($n\gtrsim10^{-3}\ \mathrm{cm}^{-3}$) while there are many density cut clouds that are below this threshold.} While further exploration of these different identification methods is warranted, throughout the rest of this section, we will explore those clouds selected via the temperature cut technique.
	
	For the temperature cut clouds, we also explore the cloud mass$-$size relation in Figure~\ref{fig:mass-size}.
	Here we are showing \rev{the kernel density estimations of}{} the distributions of cloud masses as a function of cloud size for the \rev{different}{Default and 200~pc} resolution simulations \rev{}{with power law fits shown as lines}. \rev{We have also performed power law fits to the Default and 200~pc resolution simulations.}{} For the default resolution case, we have used maximum likelihood estimation of the truncated power law since there is a cutoff at $M=5.4\times10^{5}\ \text{\msun}=10\times m_b$ where $m_b$ is the baryonic mass resolution. This is because we are only considering clouds that are resolved by at least 10 cells. However, we find that the cloud detection starts to drop off around $M=8\times10^5$~\msun which is what we use for the cutoff value for the maximum likelihood estimation. For the 200~pc resolution simulation, we use the \texttt{polyfit} function from the \texttt{numpy} Python package which implements a least squares fitting routine. With our fitting function as \rev{$M=Br^m$}{$\log M=m\log r + b$}, we find $b=5.05\pm0.19$, $m=2.22\pm0.23$, and $b=5.26\pm0.01$, $m=2.34\pm0.02$ for the Default and 200~pc resolution simulations, respectively. The normalizations are different by $1.1\sigma$ and the slopes are $0.5\sigma$ different. Thus, even though we still expect to see smaller clouds as we continue to increase the resolution past 200~pc, we can expect that they will continue to follow this mass$-$size relation until some physically motivated cutoff.
	
	With 200~pc resolution in the CGM, we are able to resolve finer cloud structures, but we are also able to better trace the cloud boundary layers to probe gas mixing. To do this, we calculate radial temperature profiles from the edge of the cloud out into the surrounding ambient CGM, giving us a sense of the properties and dynamics of the mixing boundary layer.
	Using the native Voronoi mesh, we find all the neighbors of the cells comprising a cloud and define that as the boundary layer (excluding all cells that are included in the original cloud). Repeating this several times, we gradually build out a series of boundary layers around each cloud. This method allows us to average properties over contiguous boundary layers without implicitly assuming any specific cloud geometry, as most clouds have highly irregular shapes. We then define the thickness of each boundary layer as the cube root of the average volume of all cells in that layer. Thus the distance to any given layer is the sum of all the layer thicknesses interior to the layer of interest.
	
	\begin{figure}
		\centering
		\includegraphics[width=0.8\columnwidth]{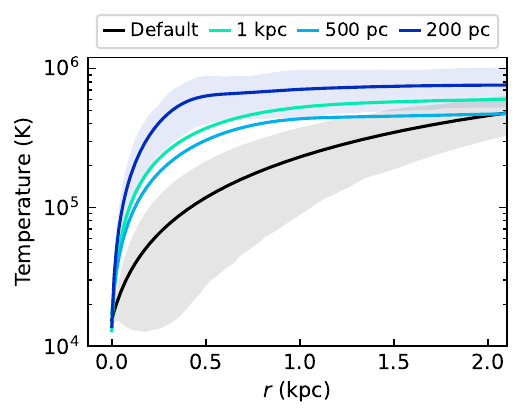}
		\caption{Radial temperature profiles of gas surrounding clouds. Each curve is constructed by taking the median temperature in each sequential boundary layer stepping out in radius from the cloud boundary. The $x$-axis is the distance from the cloud boundary \rev{in units of the cloud effective radius. The position of each boundary layer along this axis is}{} determined based on the cube root of the average volume of cells in the boundary layer. The curves are then interpolated with a cubic spline. As in Figures~\ref{fig:distributions} and \ref{fig:cloudrhot}, we are restricted to clouds within $50<r<100$~kpc\rev{}{, but also include only clouds with $1.5<r_\mathrm{eff}<4$~kpc such that we have a similar number and distribution of clouds for each simulation}. \rev{}{The 16--84\% ranges are shown for the Default and 200~pc simulations as shaded regions.} With additional resolution elements, we see a sharper transition to the ambient CGM temperature\rev{ $-$ within about half an effective radius instead of one $r_\mathrm{eff}$}{}.}
		\label{fig:radial}
	\end{figure}
	
	In Figure~\ref{fig:radial}, we show the temperature profiles for the gas surrounding the clouds for each simulation resolution. \rev{}{In order to accurately compare across simulations, we have restricted this calculation to only clouds with effective radii between 1.5 and 4~kpc, and within the CGM ($50<r<100$~kpc), as this is the region in which all four simulations overlap (see Figure~\ref{fig:distributions}).} For each cloud,\rev{ in the CGM region ($50<r<100$~kpc)}{} we calculate its radial temperature profile by taking the median temperature of all the cells in each boundary layer as a function of distance to the boundary layer in kpc. \rev{We then rescale by $r_\mathrm{eff}$ and}{We then} interpolate with a cubic spline to obtain profiles for each cloud in the given simulation. Each curve in Figure~\ref{fig:radial} is the \rev{median}{mean} of all the profiles in each simulation.
	The main result from Figure~\ref{fig:radial} shows how the temperature profile changes with increasing resolution. The slope of the temperature profile from within the cloud to the ambient medium steepens with the inclusion of volume refinement. This suggests that resolving additional cloud structure and boundary layer processes in better detail reduces the intermediate temperature layer as the clouds mix with their surroundings via diffusion or other turbulent or chaotic mixing processes. This allows clouds in the higher resolution simulations to reach the surrounding temperatures in the CGM more nearby the cloud, characterized by the sharper corners as the profiles flatten to the background temperature, suggesting that higher resolution more accurately captures the extent of gas mixing in the CGM. This was proposed in the original \textsc{tempest} paper \citep{hummels19} with their Figures~13 and 14 giving a nice graphical demonstration of the shrinking of the interface layer, \rev{but it was unconfirmed until now}{and has been seen in other recent works such as \citet{cadiou25}}.
	
	Finally, we investigate the survival of these cold clouds based on their current properties in comparison with small-scale cloud-crushing simulations \citep[e.g.,][]{kwak11,gronke18,gronke22,dutta25}. In these works, cloud survival is usually parameterized based on the cloud overdensity, Mach number, and size. Calculating these properties in our simulations, we find overdensities increasing with resolution, consistent Mach numbers across resolution, and decreasing cloud sizes as discussed above. For the overdensities, $\chi$, values range from $\chi=4-12$ for the default refinement, up to $\chi=4-25$ in the 200~pc run; the Mach numbers are $\mathcal{M}=0.2-0.5$ for all simulations; and the cloud radii systematically decrease from a couple of kpc down to $300-600$~pc in the highest resolution simulation (see Figure~\ref{fig:distributions}). Using equations (11), (24), and (27) from \citet{tan23}, we find ratios of the cloud growth times to cloud crushing times increasing slightly with increasing resolution. The 16\% to 84\% ranges for the different simulations are $t_\mathrm{grow}/t_\mathrm{cc}=0.7-1.4$, $0.8-1.8$, $1.1-2.4$, and $1.1-2.6$ for the Default, 1~kpc, 500~pc, and 200~pc resolution runs, respectively. In all simulations $>98\%$ of the clouds have $t_\mathrm{grow}/t_\mathrm{cc}$ ratios less than the fiducial value of 4 determined from the cloud crushing simulations in \citet{tan23}, indicating that the vast majority of these clouds should survive and grow.
	This is consistent with the result that the number of large clouds is converged between simulation runs (see Figure~\ref{fig:distributions}). The large clouds that are identified in the Default runs are stable and can grow \citep{lucchini24} and as the resolution is increased, the number of smaller stable clouds increases without many of the large clouds fragmenting.
	
	\begin{figure*}
		\centering
		\includegraphics[width=1.0\textwidth]{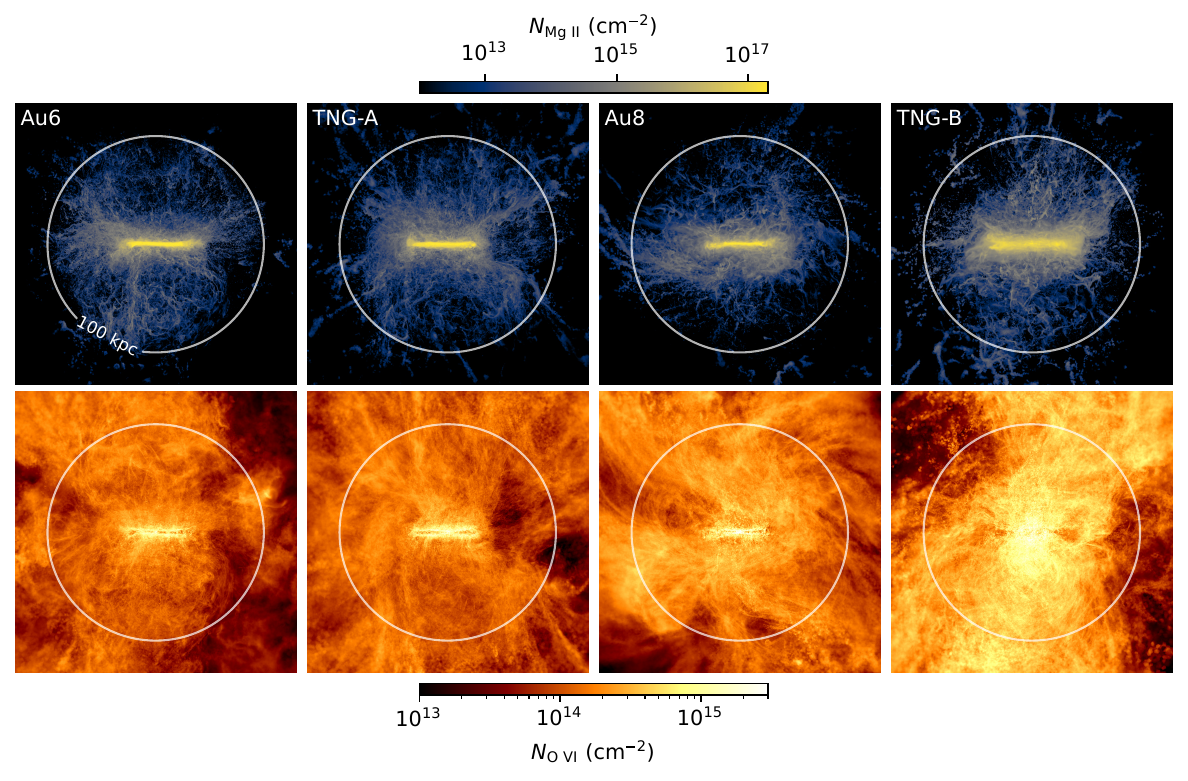}
		\caption{\mgii\ and \ovi\ column density projections for our four galaxies at $z=0$: Au6, TNG-A, Au8, and TNG-B (see Table~\ref{tab:galaxies}). Each panel shows the results from the highest resolution simulations of each galaxy: 200~pc for Au6 and TNG-A, and 500~pc for Au8 and TNG-B. As in Figure~\ref{fig:proj}, each panel is 260~kpc per side with the white circle denoting $r=100$~kpc, the limit of the fixed volume refinement region. These images were made with the ion fractions calculated with COLT RT.}
		\label{fig:4projections}
	\end{figure*}
	
	\begin{figure*}
		\centering
		\includegraphics[width=0.95\linewidth]{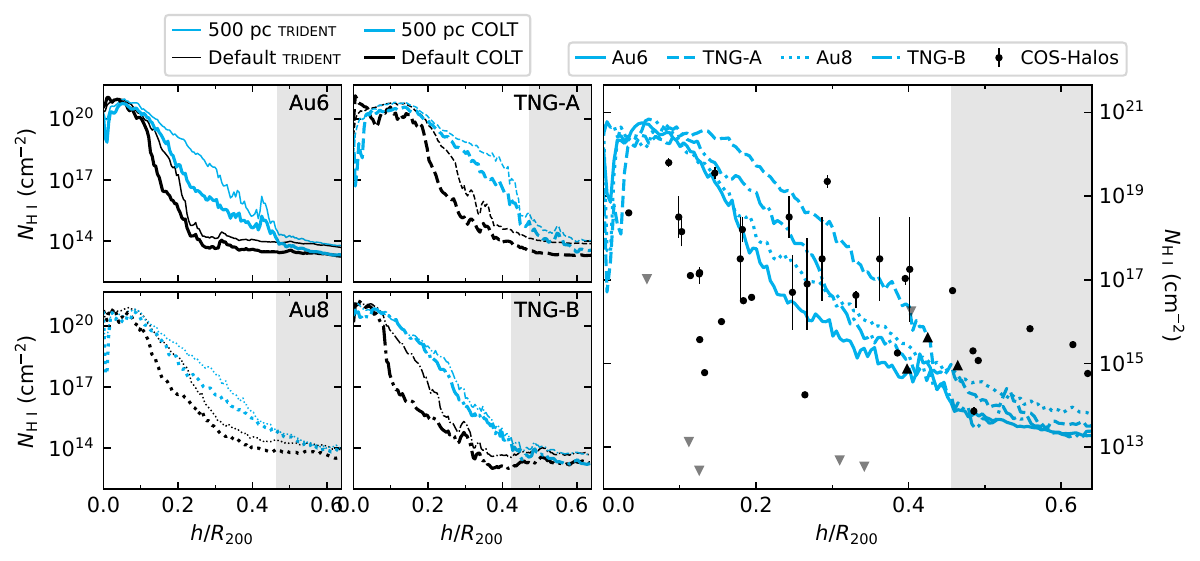}
		\caption{\hi\ radial column density profiles for our four galaxies at $z=0$ at $20^\circ$ inclination. On the left half, there is a panel for each galaxy showing the default resolution simulation in black and the 500~pc resolution simulation in light blue. The thin lines are the column density profiles using \trident\ and the thick lines are after accounting for stellar radiation with COLT. On the right all 500~pc simulation curves (calculated with COLT) are combined and overlaid are the COS-Halos \hi\ observations \citep{tumlinson13,prochaska17}. Upward facing arrows are lower limits, downward facing arrows are upper limits, and circles are measurements. The gray shaded regions denote that for $r>100$~kpc, the resolution begins to degrade. Since the virial radius of each galaxy is different, the specific location of this band is different for each panel.}
		\label{fig:4profiles}
	\end{figure*}
	
	\begin{figure*}
		\centering
		\includegraphics[width=0.95\linewidth]{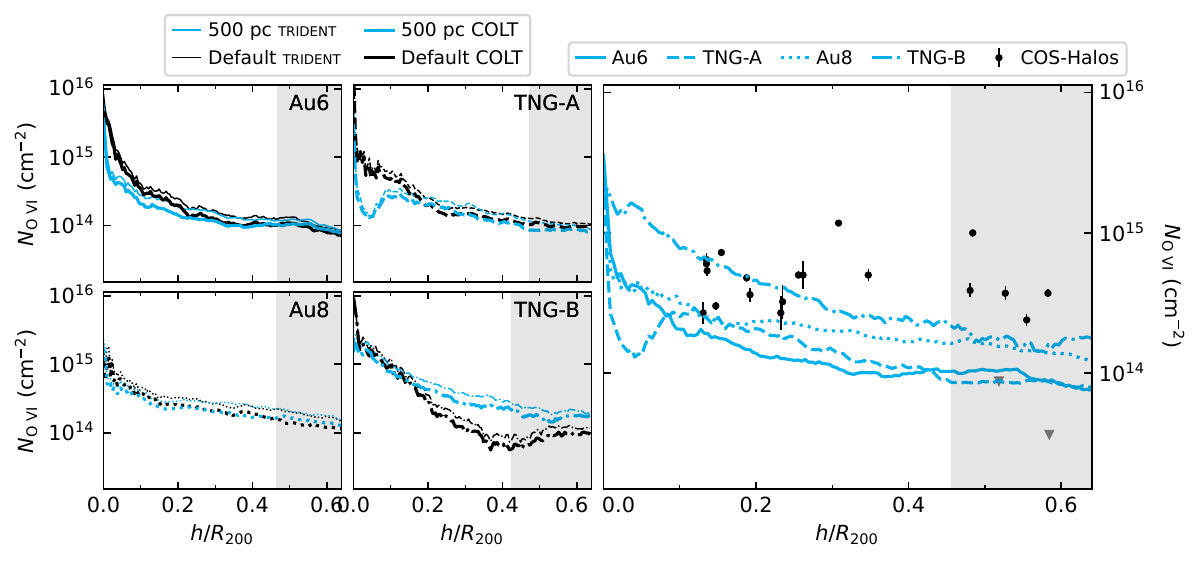}
		\caption{As Figure~\ref{fig:4profiles} except for \ovi. Observational data from \citet{tumlinson11} are included in the right panel.}
		\label{fig:4profiles-ovi}
	\end{figure*}
	
	\section{Galaxy$-$Galaxy Variation} \label{sec:4gals}
	
	The full ENGAWA suite contains cosmological zoom-in simulations for four different galaxies: two from the Auriga suite, and two selected from TNG50. We ran all four galaxies with default mass refinement, as well as fixed 1~kpc and 500~pc resolutions. In addition, Auriga halo 6 and TNG-A (ID 537941) were run at a resolution of 200~pc. All the galaxy properties are outlined in Table~\ref{tab:galaxies}.
	
	Figure~\ref{fig:4projections} shows the \mgii\ and \ovi\ column densities in projection for the four galaxies at $z=0$. Each panel shows the highest resolution fixed volume refinement simulations (200~pc for Au6 and TNG-A, and 500~pc for Au8 and TNG-B). These column densities were calculated using the COLT \rev{RT updated}{} ion fractions. Overall, we see similar trends across all galaxies: an increase in \hi\ and \mgii\ column densities in the high resolution simulations, and smoother, more filamentary \ovi. The other most striking result in this figure is the range of column densities of \ovi\ across the different galaxies. This is due to the difference in star formation histories and AGN energy injection. This is explored further in Section~\ref{sec:discussion}.
	
	Figures~\ref{fig:4profiles} and \ref{fig:4profiles-ovi} show the column densities of \hi\ and \ovi, respectively, as a function of impact parameter for the four galaxies. On the left, each panel shows the result for a single galaxy with the black lines denoting the default mass-refinement run, and the blue lines showing the result for the 500~pc volume refinement run. As in Figure~\ref{fig:colt_columns}, the thin lines are the results with \trident\ and the thick lines use COLT\rev{ RT}{}. The line styles are unique for each galaxy and can be matched up with the lines in the right panel showing all four galaxies compared against the COS-Halos results (\hi: \citealt{tumlinson13,prochaska17}; \ovi: \citealt{tumlinson11}). As in Figure~\ref{fig:coshalos}, the gray region at $h>100$~kpc indicates that that is the edge of the full refinement region. While there is still some spread in the mean radial profiles, all galaxies show an enhancement of \hi, bringing them into better agreement with the COS-Halos data. Furthermore, there is little resolution dependence on the \ovi\ profiles (with or without COLT), but the profiles do vary from galaxy to galaxy. Interestingly, we do see a minor resolution effect in TNG-B near $r\sim100$~kpc. Figure~\ref{fig:4projections} shows that this galaxy (right-most panels) has a more anisotropic \ovi\ distribution than the other galaxies. We believe this is contributing to the difference in column densities between the Default and 500~pc resolution runs seen in Figure~\ref{fig:4profiles-ovi}.

	\section{Discussion} \label{sec:discussion}
	
	Our new ENGAWA simulation suite shows that resolving the inner CGM at $\sim200$~pc profoundly reshapes the small-scale structures around the galaxy, boosting cool-phase covering fractions and clumping, while leaving bulk galaxy properties and hot-phase columns largely intact. The total \hi\ gas masses remain mostly unchanged (variations on the order of 30\%; Column 7 of Table~\ref{tab:props}) due to the dominance of the ISM. However, if we restrict ourselves to the CGM ($50<r<150$~kpc) the \hi\ gas mass is increased by factors of 3$-$6 (Column 9 of Table~\ref{tab:props}). Additionally, we see more high density gas cells as the resolution is increased. All of these factors combine to produce dramatically enhanced projected \hi\ column densities (see Figures~\ref{fig:coshalos} and \ref{fig:4profiles}). These flatter profiles are more consistent with the absorption spectroscopy observations from COS-Halos as well as other high sensitivity studies such as in \citet{das24}.
	
	Interestingly we do not see dramatic changes between the 1~kpc and 500~pc simulations in terms of \hi\ column density (Figure~\ref{fig:coshalos}). Within the $50<r<100$~kpc region, the 1~kpc simulation has an increased number of gas cells with high \hi\ mass compared to the 500~pc simulation. We believe that this is due to the fact that at 1~kpc resolution, much of the CGM is refined, however the \hi\ bearing phase consists of a combination of cells that are volume limited and mass limited. The \hi\ phase in the 500~pc resolution simulation is comprised of many more low mass cells since the CGM resolution is limited by the volume criterion. This results in the 1~kpc simulation having a boosted \hi\ column density over the Default simulation due to the additional contribution of low mass cells. The 500~pc simulation results in a comparable \hi\ column density due to the reduction in high mass cells.
	
	In contrast to the \hi, we see very consistent \ovi\ column densities across simulation resolutions (see Figure~\ref{fig:coshalos}), while the galaxy to galaxy variation is more dramatic. This is due to the different star formation and AGN histories. For the 500~pc simulations, the average star formation rates since $z=0.3$ are 1.8, 2.6, 2.3, and 4.2~\msun~yr$^{-1}$ for halos Au6, TNG-A, Au8, and TNG-B, respectively. The corresponding median \ovi\ column densities for $60<r<80$~kpc are 1.4, 1.8, 2.7, and $3.7\times10^{14}$~cm$^{-2}$, respectively. While not completely monotonic, there is definitely a trend indicating more \ovi\ column for galaxies with higher SFRs. This agrees with previous results \citep{tchernyshyov23,foggie11}. Similarly, halo TNG-B which has the highest \ovi\ column densities has the highest cumulative energy injection from its central black hole, and galaxies Au6 and TNG-A have correspondingly lower values for both. \rev{This is consistent with the idea that a flickering AGN might be responsible for boosting the \ovi\ in the CGM through photoionization \citep{oppenheimer18}.}{}
	
	We can also see that ionizing radiation from stellar sources plays a significant role in the composition of the CGM. Although \rev{}{the inclusion of stellar radiation in} post-processing \rev{RT}{} with COLT reduces the overall column densities of \hi\ around galaxies, it still increases with resolution, albeit more moderately. We believe that this is due to the fact that the higher resolution simulations produce higher density cores of \hi\ clouds in the CGM which are better protected from the radiation. However, exploring the radial column density distributions in more detail, specifically by separating the column densities seen along the major and minor axes (Figure~\ref{fig:colt_columns_major_minor}), we see that the resolution dependence is more complex. Along the major axis, the variation between the 500~pc and 200~pc resolution simulations is quite small. This indicates that in these areas of the galaxy our simulations are (or are at least close to) converged at 200~pc. Along the minor axis, however, we see that in the 50$-$100~kpc region ($0.23< h/R_\mathrm{200}< 0.47$), increased resolution still increases the observed column densities. This could be due to the post-processing nature of the RT inclusion or it could be a sign that more resolution is still required.
	
	As a way to quantify the relationship between the simulation \hi\ column density profiles and the observational data, we computed the $\chi^2$ values for the data points with respect to the simulated profiles and their 16$-84$\% ranges. For these calculations, we used the galaxies inclined at $i\sim20^\circ$ (as in Figures~\ref{fig:coshalos} and \ref{fig:colt_columns}). We define $\chi^2=\sum (N_i-N_\mathrm{sim}(r_i))^2/\sigma_\pm(r_i)$, where $N_i$ is the column density value for the $i$th observational data point, $N_\mathrm{sim}(r_i)$ is the column density for the simulation at the radial location of the $i$th observational data point, and $\sigma_\pm(r_i)$ is the difference between the median column density of the simulation at $r_i$ and the 16\% (if $N_i<N_\mathrm{sim}(r_i)$) or 84\% (if $N_i>N_\mathrm{sim}(r_i)$) limit. For this calculation, we included only measured data with uncertainties from \citet{tumlinson13} and \citet{prochaska17} from within the range $50<h<100$~kpc ($0.23< h/R_\mathrm{200}< 0.47$). \rev{Without COLT RT}{With the column densities calculated with \trident}, we see the lowest $\chi^2$ values for the 1~kpc and 500~pc resolution simulations while it increases for the 200~pc resolution simulation, indicating we are overpredicting \hi\ as the resolution continues to increase. However, \rev{after COLT RT}{with COLT}, the $\chi^2$ values consistently decrease as resolution is increased down to the the final 200~pc resolution simulation.
	
	On the other hand, accounting for stellar radiation with COLT does not dramatically impact the \mgii\ and \ovi\ profiles. However, as discussed above (see Section~\ref{sec:rt}), \rev{}{our treatment does not apply} the UVB radiation \rev{does not affect}{to} elements other than hydrogen and helium, so we are essentially seeing that the ionizing radiation from stellar sources is equivalent to the impact of the UVB when accounted for in \trident\ (without stellar radiation). Accounting for both would lead to a slight decrease in the \mgii\ columns, however we would not expect this to effect the \ovi\ profile due to its significantly higher ionization potential.
	
	Both with and without stellar radiation at all resolutions, we do see that the \ovi\ column density profiles in the simulations are slightly too low compared against the observations \citep{tumlinson11}. However, as stated above, the recent star formation history of the galaxy can play a strong role in setting the normalization of the \ovi\ columns. TNG-B, which has the highest recent star formation rate, does actually provide a good match out to $h/R_\mathrm{vir}\sim0.25$. So we expect that a certain amount of this disagreement could be due to differing recent star formation histories.
	
	\rev{}{Previous work has also shown that a flickering AGN might be responsible for boosting the \ovi\ in the CGM through photoionization \citep{oppenheimer18}. While we have not included AGN photoionization directly in the COLT analysis, we have performed some simple post-processing to understand the approximate effect that the AGN could have on the \ovi\ fraction. For all the galaxies, the BH particle have Eddington ratios of $\sim10^{-3}$, thus they are in the low-luminosity AGN (LLAGN) regime. We use an LLAGN SED \citep{nemmen11} with a luminosity calculated from a radiative efficiency scaled by the Eddington ratio, $\epsilon=0.1(\lambda_\mathrm{Edd}/0.01)$. Assuming optically thin gas, we have computed the photoionization rate (integrated across the relevant wavelengths) for each gas cell in the simulation and updated the \ovi\ ion fractions (starting with the COLT values; \citealt{verner96}). We have only considered photoionization from \ion{O}{5} into \ovi, without any exchange of \ovi\ into \ion{O}{7} (in an attempt to obtain an upper limit for the effect the AGN could have on increasing the \ovi\ columns). Within 50--100~kpc, this additional source of photoionization does not have any impact on the \ovi\ ion fractions (the ion fractions are changed by $0.3\%$). We also consider the time averaged luminosity calculated from the cumulative energy injected, which changes the \ovi\ ion fractions by up to $5\%$. Thus, we do not expect AGN photoionization to play a significant role in the \ovi\ column densities for these galaxies.}
	
	However, there could also be additional factors to consider such as the inclusion of alternative energy sources, such as cosmic rays which could change the temperature and composition of the CGM \citep{hopkins21,thomas25,ramesh25} possibly changing the fraction of oxygen that is in the \ovi\ state \citep{ji20,lu26}. \rev{}{This is discussed in more detail below.} Alternatively, the CGM metallicity in the simulations could be incorrect due to feedback yields or unphysical mixing. Finally, we are calculating the column densities in the simulation by simply summing up the densities of all the gas along the line of sight whereas in the observations, Voigt fits of absorption profiles give column density values. More precise mock observations of the simulations would illuminate whether this discrepancy could be contributing to the differences in column density values.
	
	In analyzing the properties of clouds throughout the CGM, we find a consistent mass$-$size relation across all our resolutions (Figure~\ref{fig:mass-size}). Interestingly, the slope that we determine is consistent with previous works exploring the mass$-$size relation in the ISM. While the original Larson relation \citep{larson81} found a slope of 1.9, more recent surveys using CO data have found values of 2.2$-2.3$ \citep{roman-duval10,miville-deschenes17} which are consistent with our fits: $2.22\pm0.23$ and $2.34\pm0.02$ for the Default and 200~pc resolution simulations, respectively. While a true comparison would require making mock observations in an observable ion, it is intriguing that the ISM relation is comparable to the values found in the CGM.
	
	\subsection{\rev{}{Limitations}}
	
	Some limitations of the simulations include the choices that we have made to reduce the computational cost of the simulations. We have only turned on the volume refinement scheme for $z\leq0.3$ out to 100~kpc into the halo. 3.4~Gyr have passed since $z=0.3$ which is \rev{a few}{approximately three} times the sound crossing ($R\sqrt{m_p/k_BT}\approx 1$~Gyr) and dynamical times ($R/\sigma\approx 1$~Gyr). \rev{Thus the galaxy should have had time to equilibrate after the refinement is activated,}{Therefore, if the activation of the refinement scheme introduces any large scale pressure imbalances, the galaxy would have sufficient time to come back to equilibrium.} However, its previous evolution could theoretically be altered if the refinement were to be activated for longer. Additionally, extending the refinement region out to larger radii (beyond 100~kpc) would give us a more complete picture of the outer CGM, however \rev{this cutoff should not affect results within the refined region.}{we have attempted to mitigate the effect of this cutoff on the inner CGM with the gradual change in resolution from 100 to 200~kpc (see Figure~\ref{fig:resolution}).}
	
	\rev{}{Additionally, the simulations are self-consistently tracking abundances for only 10 elements (H, He, C, N, O, Ne, Mg, Si, and Fe) with ion fractions only for \hi. Using \trident\ and COLT, we can obtain more detailed ion fractions, however both of these tools assume equilibrium and are used in post-processing, i.e. the thermodynamics of the CGM were evolved without taking the effects of local ionizing sources into account. This could affect the gas cooling and structure resulting in a different final state for the simulation. Non-equilibrium cooling effects can also play a significant role in the low density CGM leading to different ion fractions and gas properties as well. These effects are not taken into account in these simulations due to their computational cost, however with advancing technologies, such limitations may not exist in the future.}
	
	Future work will also \rev{work to}{}improve the implementation of the UVB in the COLT simulations to arise from photon sources rather than the same spatially-uniform approach taken by the simulations. We are also currently using different UVB tabulations between the simulations and the post-processing (\citealt{uvbg} vs \citealt{hm12}) however, at $z=0$ these differences only change \hi\ ion fractions by $>10$\% for $<5$\% of the cells in the CGM.
	Additionally, including the effects of the UVB radiation on more than hydrogen and helium \rev{could have a non-negligible effect on the ion fractions}{will impact the ion fractions for low-ionization species such as \mgii}.
	
	Furthermore, the inclusion of additional energy sources, such as cosmic rays, could lead to different temperatures and ion fractions. \rev{}{While the exact transport mechanisms of cosmic rays and their full effects on the CGM remain unclear, their added non-thermal support could dramatically restructure the gas around galaxies \citep{hopkins26}. Cosmic ray-driven winds are fundamentally different in nature than thermally-driven winds \citep{thomas25} and this could impact the disk-halo interface, inner CGM, as well as the global gas distribution throughout the galaxy. However, due to uncertainties in the underlying cosmic ray transport models and the lack of constraining data, their effects have not been included here. We are excited to explore the possible implications of cosmic rays at high resolution in future work.}
	
	\rev{}{Several other limitations arise from our use of the TNG galaxy formation model.} Due to the unresolved nature of the ISM, we are unable to extract details of the small-scale local structure of the disk. The TNG model has been shown to reproduce the global properties of galaxies \citep{pillepich19}, however specific star formation and feedback events are not resolved. \rev{}{Similarly, due to the implementation of the star formation and feedback model, there is a pressure floor in star forming cells corresponding to $T=10^4$~K. Therefore, we are not able to simulate any cold gas in the disk, disk-halo interface, or CGM with this model. Interestingly, previous works have not found much cold gas in the CGM, even when this temperature restriction removed \citep{stern21}; and in our simulations there are only a handful ($<10$) of individual gas cells with nonzero star formation rates in the CGM, indicating that the gas is unable to cool down much further than $10^4$~K. However, in the disk and disk-halo interface, this temperature floor plays a significant role.} In future work, we will explore the effects of improved, multiphase ISM models on the interaction between galactic disks and the inner CGM.
	\rev{Finally, only simulating four galaxies does not give a statistical sample, however it gives us a preliminary view of common trends and unique features.}{}
	
	\section{Conclusions} \label{sec:conclusions}
	
	Here we have presented the new ENGAWA (ENhanced Galactic Atmospheres With Arepo) cosmological zoom-in simulation suite including enhanced refinement in the CGM combined with the \rev{proven }{}IllustrisTNG stellar and AGN feedback model. We combine a fixed-volume refinement scheme with the default mass refinement to restrict gas cell sizes down to 200~pc within the inner 100~kpc of the CGM. This initial suite consists of four MW-like galaxies: Auriga halo 6, Auriga halo 8, TNG50 subhalo 537041, and TNG50 subhalo 519311. We run the zoom-in simulations until redshift 0.3 at which point we activate the volume refinement scheme and allow the simulation to continue for the remaining 3.4~Gyr of evolution to $z=0$. Our main conclusions are outlined below.
	\begin{enumerate}
		\item The galaxy properties (stellar and gas masses, SFRs) remain relatively unchanged with the addition of the volume refinement.
		\item Enhancing the spatial resolution within the CGM results in more, smaller clouds in the cool phase (\hi, \mgii) and we see more velocity power on small scales.
		\item Projected column densities in \hi\ and \mgii\ are enhanced by $\sim4$ orders of magnitude as the resolution is increased, bringing them into alignment with observational values.
		\item The total gas and \ovi\ column densities are not significantly changed with resolution, however the \ovi\ bearing gas is significantly more filamentary and structured at higher resolutions. Similar to previous works, the total \ovi\ column densities undercut observational values by a factor of a few.
		\item Upon inclusion of ionizing radiation from stellar sources \rev{}{(in post-processing)}, the \hi\ column densities are decreased, better aligning with observational values. The \mgii\ and \ovi\ profiles are not significantly affected by stellar radiation.
		\item The number of \rev{cold}{discrete} clouds in the CGM continues to increase as the resolution is increased. However, the number of large clouds is converged.
		\item The mass$-$size relation for the clouds is consistent with the local Larson relation with power law slopes of $\sim2.2$.
		\item The interface layers surrounding the \rev{smaller cold}{}clouds more rapidly transition to the ambient CGM gas temperature as the resolution is increased.
		\item \rev{}{The simulation snapshot files and outputs have been made publicly available\footnote{\url{https://www.scottlucchini.com/engawa}}.}
	\end{enumerate}

	\begin{acknowledgements}
		\rev{}{The authors thank the referee for their constructive comments that have improved this paper.}
		We have used initial conditions from the Auriga Project public data release \citep{grand24} available at \url{https://wwwmpa.mpa-garching.mpg.de/auriga/data}. SL thanks Ewald Puchwein for providing the TNG50 zoom ICs. SL also appreciates the valuable input from Jake Bennett, Cassi Lochhaas, and Vadim Semenov throughout the development of this project. The computations in this paper were run on the FASRC cluster supported by the FAS Division of Science Research Computing Group at Harvard University.
		Support for SL was provided by Harvard University through the Institute for Theory and Computation Fellowship.
		LH acknowledges support from the Simons Foundation through the ``Learning the Universe" initiative.
	\end{acknowledgements}
	
	\newpage
	\appendix
	
	\section{COLT Escape Fraction} \label{appendix:fesc}
	
	Due to the unresolved nature of the ISM in the TNG galaxy formation model, we have tested a variety of values for the stellar escape fraction within COLT. This parameter ($f_\mathrm{esc,*}$) reduces the emitted SED from each stellar particle with the net effect of each star particle only emitting $f_\mathrm{esc,*}$ times as many photons (applied only to frequencies above 13.6~eV). In the effective equation of state model, the cold phase (which would absorb emitted photons) is not resolved, so this parameter acts as a knob to determine how much radiation escapes the ISM.
	
	Previous studies of escape fractions from the MW ISM indicate values of $5-10$\% from superbubbles and \hii\ regions \citep{dove00} with comparable escape fractions out of the galaxies. We tested values of $f_\mathrm{esc,*}$ of 5\%, 25\%, and 100\% to evaluate the effect on the CGM ions as well as the total escape fraction at the virial radius of the galaxy. Between the fact that a value of 25\% is able to reproduce the reionization history in the THESAN simulations \citep{Kannan2022,bulichi25}, and that $f_\mathrm{esc,*}=0.25$ gives total escape fractions of ionizing photos of $5-10$\% led us to use that as our fiducial value above. However, in this appendix we show the results with alternative values of $f_\mathrm{esc,*}$.
	Table~\ref{tab:fesc} shows the escape fraction of ionizing photons at the virial radius of halo Au6 as a function of resolution and $f_\mathrm{esc,*}$.
	
	\begin{deluxetable*}{lccc}
		\tablecaption{COLT Escape Fractions}
		\label{tab:fesc}
		\tablehead{ \colhead{Resolution} & \colhead{$f_\mathrm{esc,*}=0.05$} & \colhead{$f_\mathrm{esc,*}=0.25$} & \colhead{$f_\mathrm{esc,*}=1.0$}}
		
		\startdata
		Default & 0.0041 & 0.055 & 0.37 \\
		1~kpc & 0.0056 & 0.067 & 0.39 \\
		500~pc & 0.0057 & 0.067 & 0.40 \\
		200~pc & 0.0042 & 0.054 & 0.35
		\enddata
		
		\tablecomments{Calculated by taking the escape fraction of ionizing photons ($>13.6$~eV) as output from COLT and multiplying by the $f_\mathrm{esc,*}$ parameter. Compared against $0.05-0.1$ values for the MW \citep{dove00}.}
	\end{deluxetable*}
	
\begin{figure*}
		\centering
		\includegraphics[width=1.0\textwidth]{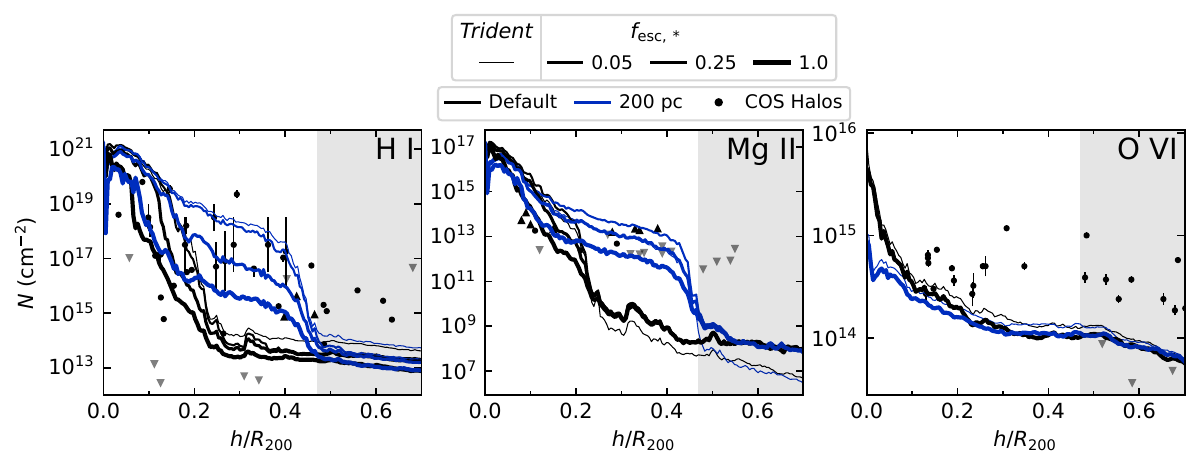}
		\caption{As Figure~\ref{fig:colt_columns} except showing the column density profiles with different choices for $f_\mathrm{esc,*}$. The black curves show the default resolution simulations while the blue lines show the 200~pc resolution simulations. The thinnest lines show the profiles calculated with \trident\ (effectively $f_\mathrm{esc,*}=0$ since there is no stellar radiation), and then increasing line thickness corresponds to increasing $f_\mathrm{esc,*}$ with values of 0.05, 0.25, and 1.0. Note that the \ovi\ profiles (and \mgii\ profiles for the default resolution beyond $h/R_\mathrm{200}=0.2$) do not change with different $f_\mathrm{esc,*}$ values, so all the COLT lines are on top of each other.}
		\label{fig:fesccomp}
	\end{figure*}

	Figure~\ref{fig:fesccomp} shows the radial profiles for \hi, \mgii, and \ovi\ for Au6 with varying values for $f_\mathrm{esc,*}$ for the Default and 200~pc resolution simulations. The thinnest lines are the profiles calculated using \trident\ (effectively the limit of $f_\mathrm{esc,*}=0$). As discussed above, the amount of \hi\ decreases with increasing radiation, and we see the same trend for \mgii. One complication with comparing the \trident\ result with the COLT outputs for \mgii\ is that \trident\ allows the UVB radiation to affect all elements and updates the ion fractions accordingly, while in COLT the UVB currently only affects hydrogen and helium. Thus we see that the \mgii\ profile from \trident\ lies on top of the $f_\mathrm{esc,*}=0.25$ line for the 200~pc resolution simulation. Interestingly, we don't see the same serendipitous behavior with the default resolution simulation, albeit the \mgii\ levels are very low to begin with.
	
	\newpage
	\bibliography{references}{}
	\bibliographystyle{aasjournalv7}
	
\end{document}